\documentclass[reprint,amsmath,amssymb,aps,floatfix,11pt]{revtex4-2}

\usepackage{graphicx}% Include figure files
\usepackage{dcolumn}% Align table columns on decimal point
\usepackage{bm}% bold math
\usepackage{physics}
\usepackage{float}
\usepackage{amsmath,amsfonts,amssymb}
\usepackage{subcaption}
\usepackage{placeins} % provides \FloatBarrier
\usepackage{hyperref}
\usepackage{wrapfig}
\usepackage[export]{adjustbox}
\usepackage{siunitx}
\usepackage{xcolor}
\usepackage{rotating}
\usepackage{breakurl}
\usepackage{geometry}
\usepackage{mathrsfs}
\begin{document}
\title{Next-Generation Heralded Single Photon Sources}
\author{Hugh Barrett$^{1,2,*}$, Imad I. Faruque$^3$}
\affiliation{$^1$Quantum Engineering Technology Labs, H. H. Wills Physics Laboratory and School of Electrical, Electronic, and Mechanical Engineering, University of Bristol, BS8 1FD, UK\\$^2$Quantum Engineering Centre for Doctoral Training, H. H. Wills Physics Laboratory and School of Electrical, Electronic, and Mechanical Engineering, University of Bristol, BS8 1FD, UK\\$^3$Department of Physics and Astronomy, University of Sheffield, Sheffield, UK\\$^*$hugh.barrett@bristol.ac.uk}
\date{\today}

\begin{abstract}

Scaling up quantum computers and building a quantum internet requires the development of ideal photon sources. Heralded single photon sources on integrated photonic platforms are the way forward to achieve this goal. Here we identify inconsistencies in source characterisation and propose methods to facilitate fairer comparison and better understanding of which sources could enable next-generation quantum applications.

\end{abstract}

\maketitle
\pagestyle{plain}
\pagenumbering{arabic}

\section{Introduction}
\label{sec:introduction}

Photons are an essential resource for emerging quantum technologies \cite{jeremy1}, with recent interest in applications including information processing \cite{bartolucci, takeda}, communications \cite{pirandola,lu}, and demonstrations of fundamental physics \cite{rauch, mohageg}. Integrated photonics in particular has risen as a promising candidate for practical and scalable applications \cite{horn}, such as quantum computers and atomic clocks with reduced size and complexity \cite{rudolph,newman}, owing to the simplicity, small physical dimensions, and stability of integrated systems \cite{matsuda}. The well-established silicon electronics industry has allowed for silicon integrated photonics to become mass-manufacturable \cite{psi}, benefitting from the abundance and CMOS compatibility of silicon \cite{silverstone2}.

The primary roadblock to using photonic platforms for quantum computing, communications, and other applications is the development of ideal single photon sources \cite{rudolph}. The ideal source emits indistinguishable photons deterministically, which is essential to perform operations and transmit information with minimal uncertainty. Indistinguishability is a key requirement for the high-visibility quantum interference necessary in linear optical quantum computing due to the DiVincenzo criteria \cite{divincenzo}. However, a single ideal source is not enough; arrays of sources are necessary to scale up information processing, with indistinguishable photons emitted from each independent source.

Heralded single photon sources (HSPSs) on integrated photonic platforms have demonstrated near-$100\%$ indistinguishability \cite{psi}, and can emit deterministically through multiplexing schemes \cite{collins}. Quantum dots are another key contender for ideal single photon sources. They have demonstrated the highest reported emission efficiency of all single photon sources \cite{ding}, but poor collection efficiencies \cite{abudayyeh,kala} prevent them from becoming fully deterministic systems \cite{li}. Furthermore, quantum dots have limited indistinguishability when scaling beyond a single source \cite{zhai}, a problem exacerbated at telecom wavelengths \cite{vajner,lio}. This makes it harder for quantum dots to utilise well-established telecommunications infrastructure, such as low-loss fibre \cite{yu}. Additionally, it is challenging to integrate quantum dots onto photonic circuits \cite{hepp}. These factors give integrated photonic HSPSs a practical edge which, alongside the advantage of reduced uncertainty in operations and information transfer, makes them the best option for demonstrating next-generation applications of quantum technology. \nocite{faruque2, faruque3,spring,llewellyn,zhu,vernon,ben1,christensen}

\begin{figure*}
    \centering
    \includegraphics[width=\textwidth]{figure0final.pdf}
    \caption{The development of next-generation HSPSs will enable next-generation applications, such as satellite-based quantum communications \cite{yin}, complex molecular simulations \cite{sparrow}, and the sensing of individual bacteria \cite{spedalieri,daher}. A timeline of milestone correlated photon pair sources are depicted. These sources are: a) Compton scattering following electron-positron annihilation \cite{bleuler}, b) bulk crystals \cite{burnham}, c) fibre \cite{bonfrate} and integrated waveguides \cite{fukuda}, d) microstructure fibre \cite{sharping}, photonic crystal waveguides \cite{suzuki}, and microring resonators \cite{clemmen}, e) photonic molecules \cite{zeng}, interferometrically coupled resonators \cite{liu}, and photonic molecule interferometrically coupled resonators \cite{integrateandscale}.}
    \label{fig:motivation}
\end{figure*}

Ideal indistinguishability requires ideal source purities such as spectral purity, P$_S$, and photon number purity \cite{faruque2, faruque3}. As a result, efforts to improve the purity of HSPSs have been explored widely \cite{spring,llewellyn,zhu,vernon,ben1,christensen}. The brightness, B, and heralding efficiency, H, of HSPSs are also critical parameters to optimise but have received comparatively less research attention. Both parameters improve scalability, making applications of HSPSs more practically realisable. This includes faster information processing \cite{kaneda,meyer-scott}, which is useful for scenarios that require a higher rate of computation protocols \cite{kaneda2} or communications over high-loss channels \cite{cao}. Higher heralding efficiency gives lower key error rates in quantum key distribution \cite{liulim} and higher rates of entanglement swapping \cite{meyer-scott2}. High brightness single photon sources are necessary for space-based quantum applications \cite{yin}, and higher photon numbers are essential for achieving quantum advantage in boson sampling \cite{neville}. Figure \ref{fig:motivation} illustrates how photon pair sources have advanced over time and provides examples of target applications.

Photonic quantum computation and communication has already been demonstrated using bulk optical components \cite{zhong, yin, steinlechner}. Integrated photonics is the natural evolution \cite{ling}, and for useful realisations of applications it is the only way forward owing to a far smaller footprint, improved scalability, and stability over bulk systems \cite{helt, matsuda}. However, most demonstrations of integrated photonic sources have low reported values of brightness compared to bulk optical equivalents, impacting their adoption. This is due to significant losses, which are caused by various factors including the use of high-loss grating couplers and parasitic processes, such as two-photon absorption (TPA) and the resulting free-carrier absorption (FCA). There are possible avenues to reduce the effects of TPA \cite{husko} and FCA \cite{engin}, and the development of low-loss couplers \cite{couplers} has resulted in count rates approaching those obtained from bulk optics \cite{du}. Nevertheless, further research into the brightness and heralding efficiency of HSPSs is essential to enable their widespread adoption for practical applications. 

In this paper we examine the state-of-the-art in integrated HSPSs. While photon counting measurements follow well-known methods, the key parameters defining source performance are inferred and their definitions are not widely agreed upon. Here we focus on the spectral purity, brightness, and heralding efficiency. Consistent definition of these parameters is crucial to measure progress and benchmark the advancement of sources for next-generation applications. For example, reporting a high value of brightness is not useful on its own; if the heralding efficiency and spectral purity are poor, then the source will not be capable of achieving next-generation applications.

\begin{figure*}
    \centering
    \includegraphics[width=\textwidth]{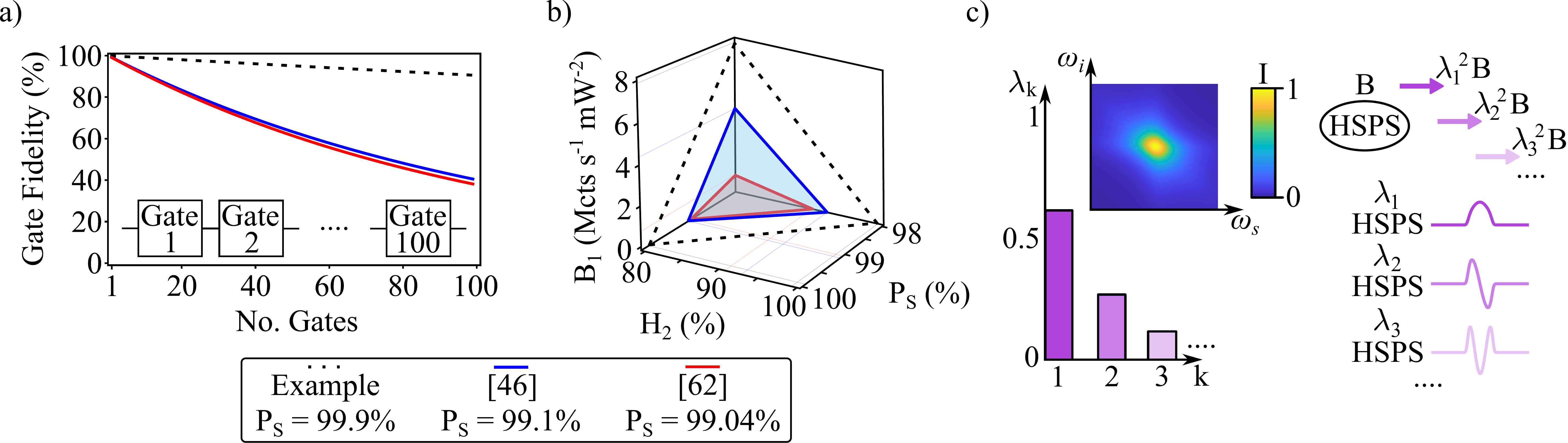}
    \caption{Attributes of HSPSs directly affecting the performance of quantum information processing. a) Imperfect spectral purity will result in exponentially decaying fidelity for subsequent gate operations. Near-$100\%$ spectral purity is therefore crucial. b) The trade-offs between brightness, heralding efficiency, and spectral purity must be overcome to optimise all parameters and achieve next-generation HSPSs for next-generation applications. High brightness and high heralding efficiency are essential for the speed of quantum operations. Values are taken from \cite{integrateandscale} and \cite{paesani}, as well as a theoretical near-ideal example. c) The fundamental relation between spectral purity and the useful brightness of a HSPS. Spectral purity, often described by the JSI (inset), can effectively be split into orthogonal Schmidt modes, $\lambda_k$. The total brightness is thus divided among the Schmidt modes according to their relative weights. Since Schmidt modes are orthogonal, they can be viewed as independent HSPSs which will not interfere in quantum information processing operations. The useful brightness is that of the fundamental Schmidt mode, which is equal to the total brightness for $100\%$ spectral purity.}
    \label{fig:optimise}
\end{figure*}

Through discussing values obtained in the literature and how they are defined, we call for consideration and clarity to facilitate the fairer comparison of HSPSs. We emphasise the need to clearly define where brightness and heralding efficiency are measured within a system, and to carefully determine the pump laser power within the generation-region. We highlight the importance of the trade-offs between parameters, and propose defining the brightness and spectral purity together to address their fundamental relationship. 

\section{Accounting for Trade-Offs}
\label{sec:trade-offs}

Recently there has been an increased focus on trade-offs between brightness, heralding efficiency, and spectral purity. Figure \ref{fig:optimise}a elaborates on why near-$100\%$ spectral purity is desirable; as indistinguishable HSPSs are required for high-visibility quantum interference \cite{thomas} to enable sufficiently high gate fidelities for quantum computation \cite{psi}. The relationship between source indistinguishability and gate fidelity varies between different types of gate \cite{micuda}, but overall fidelity will decay exponentially with an increasing number of gates \cite{wallman,gaebler}. Figure \ref{fig:optimise}a indicates the expected overall gate fidelity decay from different HSPSs, where the single gate fidelity is taken to be the spectral purity.

An example of a parameter trade-off is increasing spectral purity through narrow-band filtering, which will cause heralding efficiency to decrease \cite{integrateandscale,meyer-scott2}. For resonant structures there is a trivariate trade-off between all three parameters, for example, increasing either the spectral purity or heralding efficiency leads to a reduction in brightness \cite{lunch}. The ideal HSPS would maximise all three parameters, illustrated by Figure \ref{fig:optimise}b, so additional considerations must be made to reduce the impact of, or ideally overcome, trade-offs \cite{christensen,ben1,rodda}.

There also exists a fundamental relationship between brightness and spectral purity when practical applications are considered. The spectral purity of a HSPS is described by the joint spectral intensity (JSI) which can be plotted as an intensity profile of signal and idler frequencies, $\omega_s$ and $\omega_i$. The JSI can be decomposed into its constituent Schmidt modes, with only a HSPS with $100\%$ spectral purity emitting in just one Schmidt mode. For a non-ideal source this effectively splits the HSPS into separate HSPSs for each mode as illustrated in Figure \ref{fig:optimise}c. Each HSPS has a pure spectral mode, necessary for applications requiring indistinguishability, but with brightness split amongst them. Therefore, the useful brightness of indistinguishable photons from a HSPS with $90$\% spectral purity will be less than $90$\% of the measured brightness (this reduction being equivalent to the weighting of the fundamental Schmidt mode, $\lambda_1^2$) \cite{zielnicki}.

Imperfect photon-number purity will also reduce brightness, as the multi-pair emission effect will effectively split the HSPS into multiple HSPSs emitting different multi-photon states. However, the continued advancement and wider availability of photon number resolving (PNR) detectors will make the photon number purity less relevant \cite{moody}.

\section{Comparing HSPS\MakeLowercase{s}}
\label{sec:comparinghsps}

\subsection{Intricacies}

\begin{figure}[h]
    \centering
    \includegraphics[width=\columnwidth]{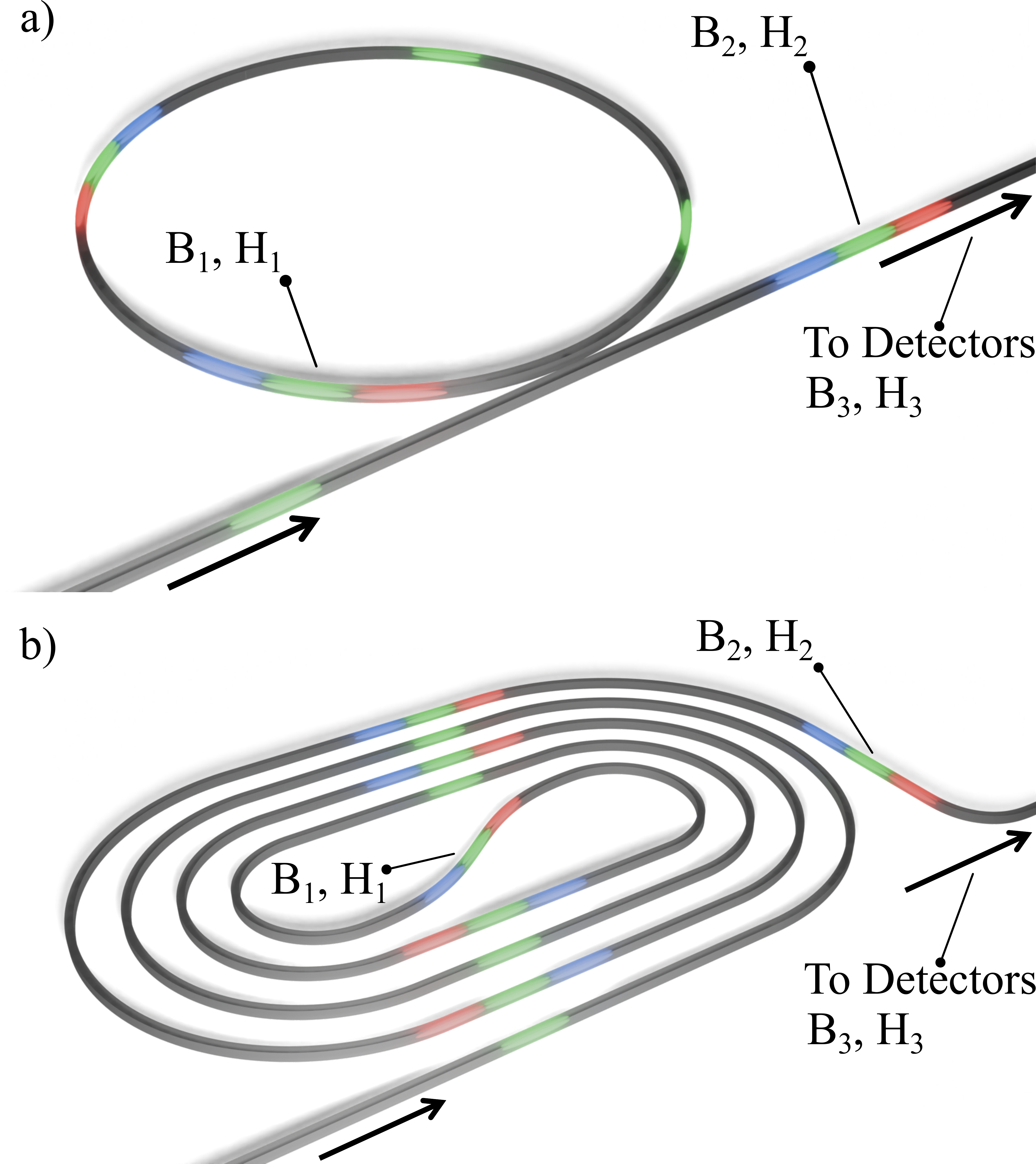}
    \caption{Defining three locations in an integrated HSPS where brightness, B, and heralding efficiency, H, can be measured for (a) a microring resonator and (b) a waveguide source.}
    \label{fig:combinedbs}
\end{figure}

Before we compare HSPSs we note that, at different points throughout a system, brightness and heralding efficiency will decrease due to photon losses. It is important to be clear with where these are defined to avoid inconsistencies that arise when comparing parameters defined in different locations \cite{integrateandscale,ma}. Similar to the definitions in \cite{esmann} but here with integrated photonics in mind, we define three distinct locations in Figure \ref{fig:combinedbs}. B$_1$ defines the effective brightness of the HSPS (also known as the intrinsic efficiency), B$_2$ is the corrected brightness just after the photon pair generation-region, and B$_3$ is the raw brightness from the coincidence count rate recorded by the detectors. Analogous to this, H$_1$ is the intrinsic heralding efficiency of the source, H$_2$ is the corrected heralding efficiency, and H$_3$ is the heralding efficiency recorded at the detectors (also known as the Klyshko efficiency). 

There are some intricacies to these definitions. H$_1$ is not a physical parameter; it is defined at the location where pair generation takes place, which could be anywhere within a ring or waveguide source, and is therefore $100\%$ by definition. Precisely where B$_1$ is defined within the generation-region is also not exact, although its value can be measured as described in Section \ref{sec:bandhdefined}. Furthermore, for waveguide sources B$_1$ and B$_2$, and similarly H$_1$ and H$_2$, both can be considered approximately equal if one ignores propagation losses due to low loss or a short generation-region. The same cannot be said for microring resonator sources due to loss mechanisms from light coupling out of the ring. These mechanisms are relatively poorly understood, leading to disagreement between theoretical and experimental results \cite{wu}. Further intricacies to the definitions of brightness and heralding efficiency are elaborated upon in Section \ref{sec:bandhdefined}. 

\subsection{Comparison}

Table \ref{table:comparison1} compares HSPSs with the most optimised parameters reported to date. A variety of source architectures are present: interferometrically coupled resonator (ICR), photonic molecule interferometrically coupled resonator (PMICR), multi-mode waveguide (MMW), and microring resonator (MRR) sources. The material platforms used for these sources are silicon (Si) and aluminium gallium arsenide (AlGaAs). Coincidence-to-accidental ratio (CAR) is listed along with the corresponding coincidence count rate ($R_{s,i}$), which can be considered as an alternative measure of the brightness output from a source. All $R_{s,i}$ values were determined on-chip by considering system losses and all sources were pumped with pulsed lasers, unless stated otherwise.

\begin{table*}
\centering
\hspace*{-0.75cm}
\begin{tabular}{|c |c |c |c |c |c |c |c |c |c|} 
\hline Reference & Year & Source & CAR & P$_S$ & H$_2$ & H$_3$ & B$_1$ & B$_2$ & B$_3$ \\
\multicolumn{1}{|c|}{} & 
\multicolumn{1}{c|}{} & 
\multicolumn{1}{c|}{} & 
\multicolumn{1}{c|}{(with $R_{s,i}$)} & 
\multicolumn{1}{c}{} & 
\multicolumn{1}{c}{\%} & 
\multicolumn{1}{c|}{} & 
\multicolumn{1}{c}{} & 
\multicolumn{1}{c}{Mcts s$^{-1}$ mW$^{-2}$} & 
\multicolumn{1}{c|}{} \\
 \hline
 \cite{psi}$^{a}$ & 2025 & Si ICR & - & 99.5 $\pm$ 0.1 & - & $\approx$ 26 & - & - & - \\
 \cite{integrateandscale}$^{b}$ & 2024 & Si PMICR & $\approx$ 305 ($\approx$ 81 kcts s$^{-1}$)  & 99.1 $\pm$ 0.1 & 93 $\pm$ 3 & 6.4 $\pm$ 0.2 & 4.4 $\pm$ 0.1 & - & - \\
 \cite{steiner}$^{c}$ & 2021 & AlGaAs MRR & 4389 (0.23 Mcts s$^{-1}$) & - & - & - & 20,000 & $\approx$ 16,800 & - \\
 \cite{paesani}$^{d}$ & 2020 & Si MMW & - & 99.04 $\pm$ 0.06 & 91 $\pm$ 9 & 12.6 $\pm$ 0.2 & 0.89 & 0.89 & 0.06 \\
 \cite{liu}$^{e}$ & 2019 & Si ICR & - & 99.1 & 52.4 & - & - & - & 2.58 \\
  \cite{ma}$^{c}$ & 2017 & Si MRR & 530 $\pm$ 40 (1.1 Mcts s$^{-1}$) & - & - & $\approx$ 3.5 &  149 $\pm$ 6 & 316 & - \\
 \cite{silverstone1}$^{f}$ & 2015 & Si MRR & $\approx$ 10 (30 cts s$^{-1}$) & 85.5 & - & - & 204 & - & 0.0013 \\ 
 [1ex] 
 \hline
\end{tabular}
\caption{Comparison between HSPSs with the most optimised parameters reported to date, dashes are given where a particular parameter was not reported. An extended comparison table can be found in the Appendix. \newline $^a$ A maximum CAR of $<$ 3000 was measured. \newline$^b$ H$_2$ and H$_3$ are the averages of signal and idler heralding efficiencies (92 ± 3 \% and 94 ± 3 \%, and 7.2 ± 0.2 \% and 5.6 ± 0.2 \% respectively). \newline$^c$ Utilised a CW, rather than a pulsed, pump laser. \newline$^d$ B$_1$ and B$_2$ are assumed equal for a waveguide source. Average pump power was measured off-chip rather than on-chip. \newline$^e$ A maximum CAR of 81 was measured. An on-chip pair generation rate of 1147 kcts s$^{-1}$ is listed at an unknown pump power. \newline$^f$ The given $R_{s,i}$ is that measured on the detectors. }
\label{table:comparison1}
\end{table*}

A CAR greater than $10$ is considered the threshold for the ability to practically resolve coincidence events \cite{savanier2}, which is achieved by all sources where the CAR was measured and therefore all are feasible for practical applications. The ICR discussed in \cite{psi} demonstrates the highest spectral purity and H$_3$, while the PMICR in \cite{integrateandscale} has the highest H$_2$ but a relatively low B$_1$. The MRR in \cite{steiner} has a significantly higher CAR, B$_1$, and B$_2$ than the other sources, while the ICR in \cite{liu} demonstrates the highest B$_3$. Paradoxically, the MMR in \cite{ma} has a higher B$_2$ than B$_1$ despite the expected photon losses, which could be due to discrepancies in the way these parameters were measured.

Greatly complicating this comparison is the large number of gaps present in Table \ref{table:comparison1} where parameters have not been reported. This makes it impossible to fairly compare HSPSs and conclude which design is best for optimising certain parameters. A wider literature search indicates that in recent years there have been efforts to report parameters more cohesively, although gaps still exist (see Appendix). The large number of footnotes necessary in Table \ref{table:comparison1} is a further indication of the current complexity of comparing HSPSs.

\begin{figure}
    \centering
    \includegraphics[width=0.85\columnwidth]{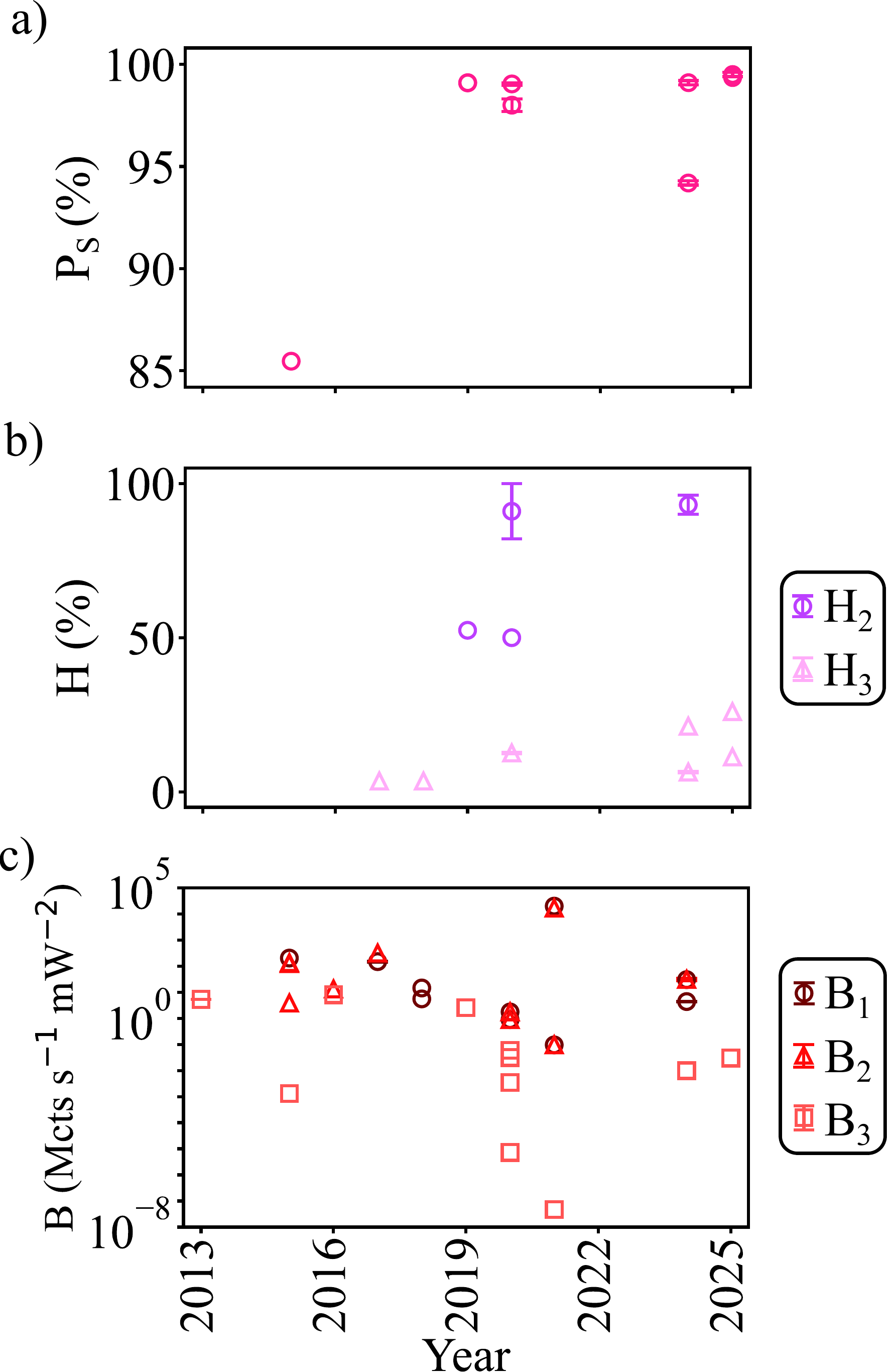}
    \caption{Reported HSPS parameters over time. The full list of references and corresponding parameter values can be found in the Appendix.}
    \label{fig:history}
\end{figure}

Although it may be difficult to compare sources, we can analyse trends in the parameters reported over time. In Figure \ref{fig:history} we see that reported spectral purity and heralding efficiency values have improved to the point of saturation, while brightness values have largely unchanged or even decreased. While this means that the indistinguishability and error rates of HSPSs have gradually improved, it is essential that attention is paid to maintaining a high brightness to ensure sources are useful for practical applications.

We cannot yet determine which source design on which material platform produces the optimal parameters for next-generation applications. Furthermore, it is apparent from Figure \ref{fig:history} that the errors on measurements are seldom considered. What is less apparent is that there exist conflicts and a lack of clarity in the way that the parameters, particularly brightness and heralding efficiency, are defined, which will be expanded upon in Section \ref{sec:considerationsforfaircomparison}. 

\section{Considerations for Fair Comparison}
\label{sec:considerationsforfaircomparison}

\subsection{Definitions}
\label{sec:bandhdefined}

Brightness does not have a strict definition. Rather, it is a measure of the photon pair generation rate and is usually modulated by parameters of the pump laser or sometimes of the source architecture. The most common modification relates to how, for sources utilising four-wave mixing, the pair generation rate is proportional to the square of the pump laser power \cite{helt}. The practical considerations here are clear for a continuous wave (CW) pump laser: simply divide the pair generation rate (in millions of counts per second, Mcts s$^{-1}$) by the average pump power squared to obtain a metric for brightness that is independent of the pump power, with units of Mcts s$^{-1}$ mW$^{-2}$. This is less straightforward when a pulsed laser is used, where one must consider the envelope function of the pulse and the repetition rate. The pair generation rate for a pulsed source is understood to be proportional to the square of the total pulse energy \cite{lunch, bonneau}. However, this is challenging to measure practically because it requires a good understanding of the temporal profile of pulses, which are usually on the scale of picoseconds and therefore can only be resolved with very accurate and costly oscilloscopes. While considering the total pulse energy would be ideal, to our knowledge, this has not yet been attempted other than an indirect approach from \cite{integrateandscale}.

For pulsed lasers, it is instead common to consider either peak or average pulse power, with the assumption that the pulse has a square-wave temporal profile. Using peak power is often justified by how the fundamental equations describing pair generation depend on field amplitudes \cite{agrawal}. However, modulating brightness in this way does not take into account varying pulse widths or repetition rates. It is also non-trivial to measure the peak power of a pulsed laser; again a high resolution oscilloscope is required. Conversely, the average pulse power is a direct observable and takes into account the pulse width and repetition rate, thus allowing an accurate comparison of the relative brightness between different sources. For these reasons, it is most common to find the brightness defined in terms of the average power \cite{paesani,silverstone1,savanier,ma,llewellyn}, a convention that we also adopt here. In no situation is it possible to draw a clear comparison between CW and pulse pumped sources, necessitating the pump regime to be clearly defined when comparing HSPSs.

There have been efforts to further adjust the definition of brightness, such as dividing by the linewidth of the pump profile for MRR, or equivalently by the bandwidth for waveguide sources \cite{jiang,bohan}. Here we consider the brightness to be the photon pair generation rate per unit squared average pump power, and the heralding efficiency to be the probability of one photon in a pair being present in the system, rather than lost to the environment, given that the other photon has been detected. Further modifications to either parameter only complicate the already arduous task of comparing different HSPSs. It is common to see brightness defined with units of MHz mW$^{-2}$. We use Mcts s$^{-1}$ mW$^{-2}$ to avoid confusing count rate with linewidth or bandwidth frequencies, and to reflect the counting nature of single photon detectors. There are also examples of entirely new figures of merit that combine several parameters \cite{husko,integrateandscale}. While these could simplify comparisons in the ideal case, they are restricted by a lack of consistency in how fundamental parameters are often defined.

Finally, it is important to consider that there are two distinct ways to measure brightness and heralding efficiency. The most straightforward method is to measure the raw value for $R_{s,i}$ and divide by the average pump power squared to calculate B$_3$, or alternatively divide by the count rate of the heralded (conventionally the idler) photons to calculate the heralding efficiency of the system as a whole \cite{klyshko1,klyshko2}. The latter is conceptually similar to H$_3$ but the two parameters cannot be equated. This is because the overall system heralding efficiency will also include dark counts, accidentals (counts from coincidence events within the coincidence window that do not correspond to generated pairs), and counts from spurious photon pair generation in nonlinear regions before and after the source itself \cite{silverstone3,benarxiv}.

Alternatively, B$_1$ (sometimes referred to as $\gamma_{eff}$ in this context) and H$_3$ for signal and idler channels can be determined through curve fitting. By considering the dependence of $R_{s,i}$ and of the singles count rates for signal, $R_{s}$, and idler, $R_{i}$, photons on the average pump power, $P_{avg}$, we can fit to Equations \ref{eqn:brightness1},\ref{eqn:brightness2}, \ref{eqn:brightness3}, and \ref{eqn:brightness4} \cite{bonneau}.

\begin{equation}
    R_{s} = H_{3,s}(B_1P_{avg}^{2} + \beta_{s}P_{avg}) + R_{DC,s}
    \label{eqn:brightness1}
\end{equation}

\vspace{-0.5cm}

\begin{equation}
    R_{i} = H_{3,i}(B_1P_{avg}^{2} + \beta_{i}P_{avg}) + R_{DC,i}
    \label{eqn:brightness2}
\end{equation}

\vspace{-0.5cm}

\begin{equation}
    R_{s,i} = H_{3,i}H_{3,s}B_1P_{avg}^{2} + R_{ACC}
    \label{eqn:brightness3}
\end{equation}

\vspace{-0.5cm}

\begin{equation}
    R_{ACC} = R_{s}R_{i}\tau
    \label{eqn:brightness4}
\end{equation}

Here $\beta$ represents linear noise terms and $R_{DC}$ the rate of dark counts, each with signal and idler channel counterparts. $R_{ACC}$ is the rate of accidentals, and $\tau$ is the width of the coincidence window \cite{faruque2}. This method carries less inherent uncertainty, as fitting can be performed over many data points, and separates the brightness and heralding efficiency of the source from that of the overall system. However, despite these benefits, heralding efficiency has only been determined in this way in \cite{psi} and \cite{integrateandscale}.

\subsection{Measurement Point}
\label{sec:measurementpoint}

When comparing different HSPSs, H$_2$ is the most valuable definition of heralding efficiency because it disregards losses from all components other than the source itself, while H$_3$ is more useful when considering the performance of the system as a whole. H$_2$ can be estimated from H$_3$ by subtracting sources of loss within the system. 

Similarly, B$_2$ is in theory the most valuable definition of brightness when comparing HSPSs because it isolates the performance of the source from the performance of other components. Determining B$_2$ through B$_3$ involves several assumptions when back-propagating as this relies on understanding all sources of loss, including non-linear loss such as TPA. Measuring this loss is non-trivial and relies on trusting that theoretical equations accurately reflect actual system behaviour. It is also possible to calculate B$_2$ by forward-propagation from a direct measurement of B$_1$, which involves less inherent uncertainty if the curve fitting is performed over a large number of data points. However, this approach carries the assumption that equations \ref{eqn:brightness1}, \ref{eqn:brightness2}, \ref{eqn:brightness3}, and \ref{eqn:brightness4} are accurate, and for resonant sources relies on measuring the poorly understood loss mechanisms from coupling out of rings. Additionally, for resonant sources this does not take different coupling regimes into consideration, as while B$_1$ is optimised by operating in the critical coupling regime, B$_2$ is optimised in a somewhat over-coupled regime \cite{lunch}.

Ideally, B$_2$ should be determined using both the forward- and back-propagation techniques and a comparison made to verify the results, similar to measuring the spectral purity via both JSI and unheralded $g^{(2)}$ methods \cite{paesani,liu,integrateandscale}. To our knowledge, the forward-propagation technique has not yet been attempted. As a result, we consider B$_1$ to currently be the best parameter to compare HSPS brightness due to the improved accuracy of its measurement; restricting B$_2$ to be the most useful for estimating the source output, and B$_3$ for considering the output from the system as a whole.

\subsection{Pump Considerations}

\begin{figure}
    \centering
    \includegraphics[width=\columnwidth]{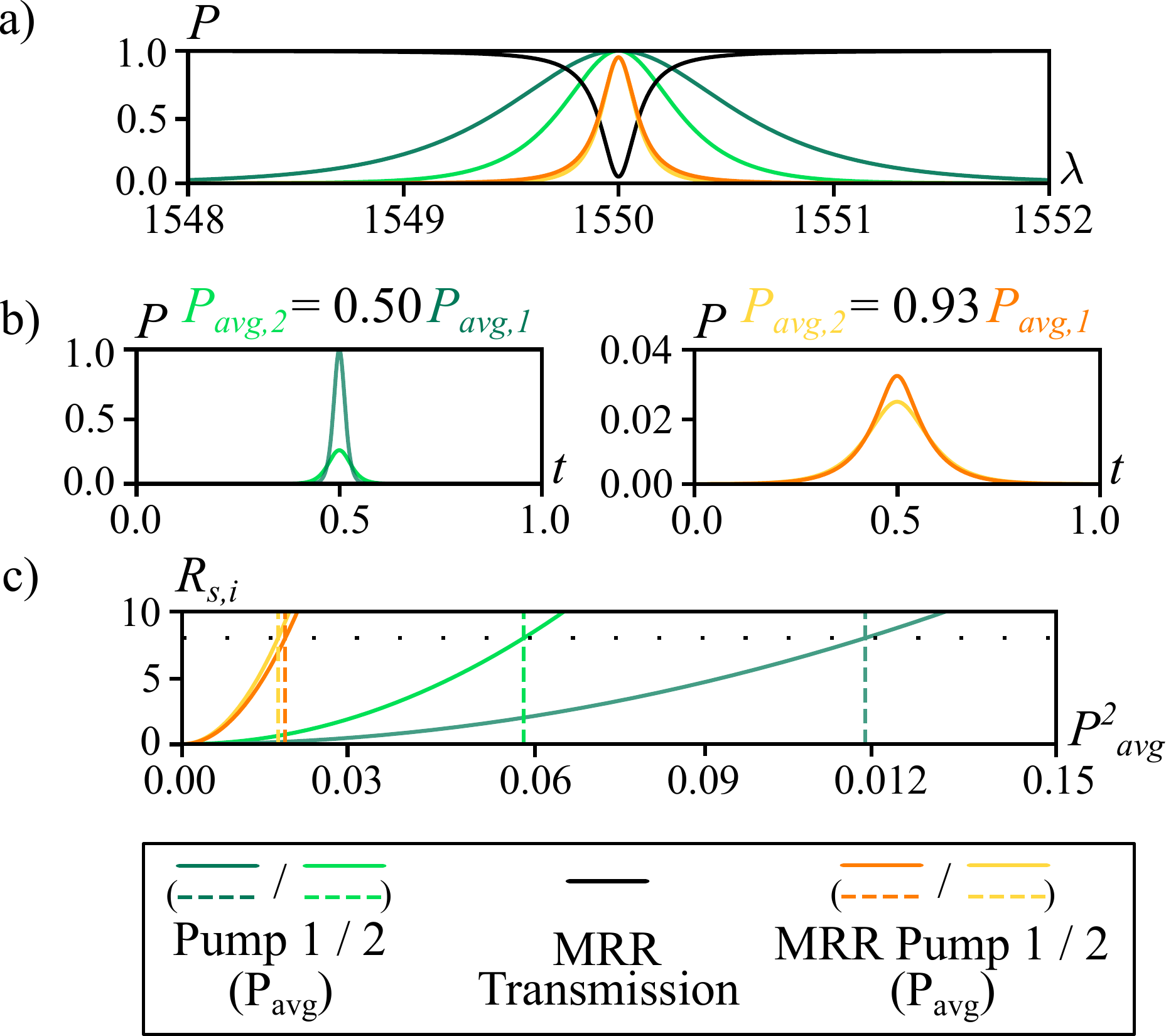}
    \caption{a) Simulated spectral profiles for two different pulse linewidths within a side-coupled waveguide, the overlap of these with the MRR transmission function gives the spectral profiles within the MRR. b) Temporal profiles corresponding to the spectral profiles, and the relative average pump powers between them. c) Simulated plots of $R_{s,i}$ against $P^2_{avg}$, where a given value of $R_{s,i}$ (indicated by the black dotted line) is measured at the average power of each pump profile.}
    \label{fig:MRRpump}
\end{figure}

Theoretical models of sources are known to often overestimate brightness when compared to experimental measurements \cite{savanier}. For MRR it is convention to measure the average pump power within the side-coupled waveguide rather than the resonant cavity itself, despite the resonator being where pair generation predominantly takes place. This is not a concern for waveguide sources, as the phase matching condition (unless very narrow) results in the majority of pump power being used to produce photon pairs. However, for MRR the phase matching condition restricts the resonator to only absorb a fraction of the pump light and, therefore, only a fraction of the average pump power from the laser is used to produce photon pairs. As such, the transmission profile of the MRR partially governs the source brightness, which could be a significant contributing factor to discrepancies between theory and experiment.

Figure \ref{fig:MRRpump}a illustrates the spectral profiles of two different pulsed laser pumps within a side-coupled waveguide, one with half the linewidth of the other. The transmission function of an MRR is also shown, and the spectral profiles of the two pumps within the MRR are determined from the overlap between this transmission function and the side-coupled waveguide profiles. In Figure \ref{fig:MRRpump}b the corresponding temporal profiles are shown, from which the average pump powers are calculated assuming a fixed repetition rate. As expected, the pump profile with half the linewidth has half the average pump power within the side-coupled waveguide; however, this change is much less significant when considering the pumps within the MRR. Finally, Figure \ref{fig:MRRpump}c simulates how measurements of $R_{s,i}$ against $P^2_{avg}$ would vary based on which pump profile is used to determine the average power. The choice of pump profile has a significant impact on the trend of the plot and hence the brightness calculated when curve fitting using equation \ref{eqn:brightness3}.

Considering the average pump pulse power within the MRR gives an increased brightness and is less sensitive to pulse linewidth. To our knowledge this consideration has not yet been made, an oversight in how brightness is characterised for resonant sources. By considering the average pump power within the resonator there is potential for closer agreement between experimental results and theoretical predictions, applicable for both CW and pulsed pump regimes. For pulsed regimes, this also facilitates fairer comparison between sources with different pulse linewidths.

However, this is not a perfect solution as the pump field within the ring will be amplified due to resonant field enhancement \cite{martini}. The power within the resonator has a time-dependent build up and exponential decay, making determining the true average power non-trivial. To account for this, a thorough theoretical model of the time dependence of photon pair generation is required.

\section{Outlook}

We have analysed the state-of-the-art in HSPS research and called for consideration into how parameters can be more clearly defined to facilitate fairer comparison between sources. Currently, it is not feasible to draw conclusions on the ideal HSPS design for a given application due to a lack of reporting on certain parameters and inconsistencies in how parameters are defined. Care should be taken to clearly define where brightness and heralding efficiency are measured within a system, and to define these parameters consistently and accurately. For pulse pumped sources, the definition of the pulse in terms of average power, or ideally the total energy should be detailed, including these values within the ring in the case of resonant sources. Finally, it is essential to consider trade-offs that exist between HSPS parameters, and how the parameters may influence each other when a particular practical application is considered.

\section{Acknowledgements}
\label{sec:acknowledgements}

H.B. thanks Jorge Barreto, Siddharth Joshi, Alex Clark, and Krishna Balram for useful discussions. The authors acknowledge funding and support from the Engineering and Physical Science Research Council (EPSRC) Hibbert Scholarship, EP/SO23607/1, and the EPSRC Quantum Communications Hubs, EP/M013472/1 and EP/T001011/1.

\section{Appendix}
\label{sec:appendix}

\begin{sidewaystable*}[h]
\centering
\vspace*{9cm}
\begin{tabular}{|c |c |c |c |c |c |c |c |c |c |c|} 
 \hline Reference & Year & Source & CAR & P$_S$ & P$_n$ & H$_2$ & H$_3$ & B$_1$ & B$_2$ & B$_3$  \\
\multicolumn{1}{|c|}{} & 
\multicolumn{1}{c|}{} & 
\multicolumn{1}{c|}{} & 
\multicolumn{1}{c|}{(with $R_{s,i}$)} & 
\multicolumn{1}{c}{} & 
\multicolumn{1}{c}{} & 
\multicolumn{1}{c}{\hspace*{-1.8cm}\%} & 
\multicolumn{1}{c|}{} & 
\multicolumn{1}{c}{} & 
\multicolumn{1}{c}{Mcts s$^{-1}$ mW$^{-2}$} & 
\multicolumn{1}{c|}{} \\
 \hline
 \cite{psi}$^{a\hphantom{aa}}$ & 2025 & Si ICR & - & 99.5 $\pm$ 0.1 & 99.64 $\pm$ 0.02 & - & $\approx$ 26 & - & - & -\\
 \cite{psi}$^{\hphantom{aaa}}$ & 2025 & Si CMRR & - & 99.35 & - & - & - & - & - & -\\
 \cite{wang}$^{b\hphantom{aa}}$ & 2025 & Si SMW & 16.77 $\pm$ 0.01 (1.51 Mcts s$^{-1}$) & - & 99.906 $\pm$ 0.002 & - & 11.3 & - & - & 0.03\\
 \cite{integrateandscale}$^{c\hphantom{aa}}$ & 2024 & Si PMICR & $\approx$ 305 ($\approx$ 81 kcts s$^{-1}$) & 99.1 $\pm$ 0.1 & 99.7 $\pm$ 0.2 & 93.1 $\pm$ 3.1 & 6.4 $\pm$ 0.2 & 4.4 $\pm$ 0.1 & - & -\\
 \cite{du}$^{b,d\hphantom{a}}$ & 2024 & Si SMW & 27 (1.21 Mcts s$^{-1}$) & - & 99.96 & - & 21.3 & - & - & 0.00965\\
 \cite{rahmouni}$^{d,e\hphantom{a}}$ & 2024 & SiC MRR & 620 (9 $\pm$ 1 kcts s$^{-1}$) & - & $>$ 99.9 & - & - & - & 31 $\pm$ 3 & -\\
 \cite{zhu2}$^{b,f\hphantom{a}}$ & 2024 & SiN SMW & 149 ($\approx$ 0.01 kcts s$^{-1}$) & - & - & 29 & 18 & - & - & $\approx$ 0.01\\
 \cite{chen}$^{b,d,g}$ & 2024 & SiN MRR & 1438 $\pm$ 22 (22 cts s$^{-1}$) & 94.2 $\pm$ 0.1 & 99.63 $\pm$ 0.05 & - & - & 30.4 & - & -\\
  \cite{steiner}$^{d\hphantom{aa}}$ & 2021 & AlGaAs MRR & - & - & 100 $\pm$ 1 & - & - & 20,000 & $\approx$ 16,800 & - \\
  \cite{mahmudlu}$^{d,h\hphantom{a}}$ & 2021 & AlGaAs SMW & 4389 (0.23 Mcts s$^{-1}$) & - & - & - & - & 0.096 & 0.096 & 4.7$\times10^{-8}$ \\
 \cite{ben1}$^{\hphantom{aaa}}$ & 2020 & Si MRR & - & 98.0 $\pm$ 0.3 & - & - & - & - & - & 0.00345\\
 \cite{paesani}$^{f,h}$ & 2020 & Si MMW & - & 99.04 $\pm$ 0.06 & $\approx$ 99.9 & 91 $\pm$ 9& 12.6 $\pm$ 0.2 & 0.89 & 0.89 & 0.06\\
 \cite{llewellyn}$^{b\hphantom{aa}}$ & 2020 & Si MRR & $>$ 50 ($\approx$ 20 kcts s$^{-1}$) & - & $\approx$ 98.9 & 50 & - & - & - & 0.03125\\
 \cite{choi}$^{h\hphantom{aa}}$ & 2020 & USRN SMW & $\approx$ 3 (0.44 Mcts s$^{-1}$) & - & - & - & - & - & 1.76 & 7.2$\times10^{-6}$\\
 \cite{liu}$^{i\hphantom{aa}}$ & 2019 & Si ICR & - & 99.1 & - & 52.4 & - & - & - & 2.58\\
 \cite{faruque1}$^{j\hphantom{aa}}$ & 2018 & Si MRR & - & - & - & - & - & 5.572 & - & -\\
 \cite{ma2}$^{k\hphantom{aa}}$ & 2018 & Si MRR & - & - & 99.5 & - & $\approx$ 3.5 & 14.6 & - & -\\
 \cite{ma}$^{d\hphantom{aa}}$ & 2017 & Si MRR & 530 $\pm$ 40 (1.1 Mcts s$^{-1}$) & - & 99 $\pm$ 2 & - & $\approx$ 3.5 & 149 $\pm$ 6 & 316 & -\\
 \cite{savanier}$^{\hphantom{aaa}}$ & 2016 & Si MRR & 65 (83 kcts s$^{-1}$) & - & - & - & - & - & 13.3 & -\\
 \cite{mazeas}$^{d\hphantom{aa}}$ & 2016 & Si MRR & - & - & - & - & - & - & - & 8\\
 \cite{silverstone1}$^{b\hphantom{aa}}$ & 2015 & Si MRR & 10 (30 cts s$^{-1}$) & 85.47 & - & - & - & 204 & - & 0.0013\\
 \cite{savanier2}$^{d\hphantom{aa}}$ & 2015 & Si MRR & $\approx$ 21 ($\approx$ 37 kcts s$^{-1}$) & - & - & - & - & - & 3.84 & -\\
 \cite{wakabayashi}$^{d\hphantom{aa}}$ & 2015 & Si MRR & 350 $\pm$ 10 (21 Mcts s$^{-1}$)& - & - & - & - & - & 124.93 & -\\
 \cite{jiang}$^{d\hphantom{aa}}$ & 2015 & Si MD & 270 $\pm$ 10 (855 kcts s$^{-1}$)& - & - & - & - & - & 136.98 & -\\
 \cite{engin}$^{b,d\hphantom{a}}$ & 2013 & Si MRR & 37 $\pm$ 1 (123 Mcts s$^{-1}$ $\pm$ 11 kcts s$^{-1}$)& - & - & - & - & - & - & 5.3385 $\pm$ 0.0005\\
 [1ex] 
 \hline
\end{tabular}
\caption{Comparison between HSPS parameters, dashes are given where a particular parameter was not reported. The types of material platform are: silicon (Si), silicon carbide (SiC), silicon nitride (SiN), aluminium gallium arsenide (AlGaAs), and ultra-silicon-rich nitride (USRN). The types of source architectures are: interferometrically coupled resonator (ICR), cascaded microring resonator (CMRR), single-mode waveguide (SMW), photonic molecule interferometrically coupled resonator (PMICR), microring resonator (MRR), multi-mode waveguide (MMW), and microdisk (MD) sources. Listed parameters are: coincidence-to-accidental ratio (CAR) along with the corresponding coincidence count rate ($R_{s,i}$), spectral purity (P$_S$), maximum photon number purity (P$_n$), heralding efficiency (H), and brightness (B). \newline $^{a}$ A maximum CAR of $<$ 3000 was measured. \newline$^b$ The given $R_{s,i}$ is that measured on the detectors. \newline $^c$ H$_2$ and H$_3$ are the averages of signal and idler heralding efficiencies (92 ± 3 \% and 94 ± 3 \%, and 7.2 ± 0.2 \% and 5.6 ± 0.2 \% respectively) \newline$^d$ Utilised a CW, rather than a pulsed, pump laser. \newline $^{e}$ A maximum CAR of 21 was measured. \newline$^f$ Average pump power was measured off-chip rather than on-chip. \newline$^g$ Spectral purity was determined using an unheralded g$^{(2)}$ measurement rather than the JSI. \newline$^h$ B$_1$ and B$_2$ are assumed equal for a waveguide source. \newline$^i$ A maximum CAR of 81 was measured. An on-chip pair generation rate of 1147 kcts s$^{-1}$ is listed at an unknown pump power. \newline$^j$ B$_1$ is averaged over two different sources, with 5.013 Mcts s$^{-1}$ mW$^{-2}$ and 6.130 Mcts s$^{-1}$ mW$^{-2}$.\newline $^{k}$ A maximum CAR of 3000 $\pm$ 500 was measured.}
\label{table:comparison}
\end{sidewaystable*}

\clearpage

\bibliography{references}

\begin{thebibliography}{96}%
\makeatletter
\providecommand \@ifxundefined [1]{%
 \@ifx{#1\undefined}
}%
\providecommand \@ifnum [1]{%
 \ifnum #1\expandafter \@firstoftwo
 \else \expandafter \@secondoftwo
 \fi
}%
\providecommand \@ifx [1]{%
 \ifx #1\expandafter \@firstoftwo
 \else \expandafter \@secondoftwo
 \fi
}%
\providecommand \natexlab [1]{#1}%
\providecommand \enquote  [1]{``#1''}%
\providecommand \bibnamefont  [1]{#1}%
\providecommand \bibfnamefont [1]{#1}%
\providecommand \citenamefont [1]{#1}%
\providecommand \href@noop [0]{\@secondoftwo}%
\providecommand \href [0]{\begingroup \@sanitize@url \@href}%
\providecommand \@href[1]{\@@startlink{#1}\@@href}%
\providecommand \@@href[1]{\endgroup#1\@@endlink}%
\providecommand \@sanitize@url [0]{\catcode `\\12\catcode `\$12\catcode `\&12\catcode `\#12\catcode `\^12\catcode `\_12\catcode `\%12\relax}%
\providecommand \@@startlink[1]{}%
\providecommand \@@endlink[0]{}%
\providecommand \url  [0]{\begingroup\@sanitize@url \@url }%
\providecommand \@url [1]{\endgroup\@href {#1}{\urlprefix }}%
\providecommand \urlprefix  [0]{URL }%
\providecommand \Eprint [0]{\href }%
\providecommand \doibase [0]{https://doi.org/}%
\providecommand \selectlanguage [0]{\@gobble}%
\providecommand \bibinfo  [0]{\@secondoftwo}%
\providecommand \bibfield  [0]{\@secondoftwo}%
\providecommand \translation [1]{[#1]}%
\providecommand \BibitemOpen [0]{}%
\providecommand \bibitemStop [0]{}%
\providecommand \bibitemNoStop [0]{.\EOS\space}%
\providecommand \EOS [0]{\spacefactor3000\relax}%
\providecommand \BibitemShut  [1]{\csname bibitem#1\endcsname}%
\let\auto@bib@innerbib\@empty
%</preamble>
\bibitem [{\citenamefont {O'Brien}\ \emph {et~al.}(2009)\citenamefont {O'Brien}, \citenamefont {Furusawa},\ and\ \citenamefont {Vučkovic}}]{jeremy1}%
  \BibitemOpen
  \bibfield  {author} {\bibinfo {author} {\bibfnamefont {J.}~\bibnamefont {O'Brien}}, \bibinfo {author} {\bibfnamefont {A.}~\bibnamefont {Furusawa}},\ and\ \bibinfo {author} {\bibfnamefont {J.}~\bibnamefont {Vučkovic}},\ }\bibfield  {title} {\bibinfo {title} {{Photonic quantum technologies}},\ }\href@noop {} {\bibfield  {journal} {\bibinfo  {journal} {Nat. Photon.}\ }\textbf {\bibinfo {volume} {3}} (\bibinfo {year} {2009})}\BibitemShut {NoStop}%
\bibitem [{\citenamefont {Bartolucci}\ \emph {et~al.}(2023)\citenamefont {Bartolucci} \emph {et~al.}}]{bartolucci}%
  \BibitemOpen
  \bibfield  {author} {\bibinfo {author} {\bibfnamefont {S.}~\bibnamefont {Bartolucci}} \emph {et~al.},\ }\bibfield  {title} {\bibinfo {title} {{Fusion-based quantum computation}},\ }\href@noop {} {\bibfield  {journal} {\bibinfo  {journal} {Nat. Commun.}\ }\textbf {\bibinfo {volume} {14}} (\bibinfo {year} {2023})}\BibitemShut {NoStop}%
\bibitem [{\citenamefont {Takeda}\ and\ \citenamefont {Furusawa}(2019)}]{takeda}%
  \BibitemOpen
  \bibfield  {author} {\bibinfo {author} {\bibfnamefont {S.}~\bibnamefont {Takeda}}\ and\ \bibinfo {author} {\bibfnamefont {A.}~\bibnamefont {Furusawa}},\ }\bibfield  {title} {\bibinfo {title} {{Toward large-scale fault-tolerant universal photonic quantum computing}},\ }\href@noop {} {\bibfield  {journal} {\bibinfo  {journal} {APL Photon.}\ }\textbf {\bibinfo {volume} {4}} (\bibinfo {year} {2019})}\BibitemShut {NoStop}%
\bibitem [{\citenamefont {Pirandola}\ \emph {et~al.}(2020)\citenamefont {Pirandola} \emph {et~al.}}]{pirandola}%
  \BibitemOpen
  \bibfield  {author} {\bibinfo {author} {\bibfnamefont {S.}~\bibnamefont {Pirandola}} \emph {et~al.},\ }\bibfield  {title} {\bibinfo {title} {{Advances in Quantum Cryptography}},\ }\href@noop {} {\bibfield  {journal} {\bibinfo  {journal} {Adv. Opt. Photon.}\ }\textbf {\bibinfo {volume} {12}} (\bibinfo {year} {2020})}\BibitemShut {NoStop}%
\bibitem [{\citenamefont {Lu}\ \emph {et~al.}(2022)\citenamefont {Lu} \emph {et~al.}}]{lu}%
  \BibitemOpen
  \bibfield  {author} {\bibinfo {author} {\bibfnamefont {C.}~\bibnamefont {Lu}} \emph {et~al.},\ }\bibfield  {title} {\bibinfo {title} {{Micius quantum experiments in space}},\ }\href@noop {} {\bibfield  {journal} {\bibinfo  {journal} {Rev. Mod. Phys.}\ }\textbf {\bibinfo {volume} {94}} (\bibinfo {year} {2022})}\BibitemShut {NoStop}%
\bibitem [{\citenamefont {Rauch}\ \emph {et~al.}(2018)\citenamefont {Rauch} \emph {et~al.}}]{rauch}%
  \BibitemOpen
  \bibfield  {author} {\bibinfo {author} {\bibfnamefont {D.}~\bibnamefont {Rauch}} \emph {et~al.},\ }\bibfield  {title} {\bibinfo {title} {{Cosmic Bell Test Using Random Measurement Settings from High-Redshift Quasars}},\ }\href@noop {} {\bibfield  {journal} {\bibinfo  {journal} {Phys. Rev. Lett.}\ }\textbf {\bibinfo {volume} {121}} (\bibinfo {year} {2018})}\BibitemShut {NoStop}%
\bibitem [{\citenamefont {Mohageg}\ \emph {et~al.}(2022)\citenamefont {Mohageg} \emph {et~al.}}]{mohageg}%
  \BibitemOpen
  \bibfield  {author} {\bibinfo {author} {\bibfnamefont {M.}~\bibnamefont {Mohageg}} \emph {et~al.},\ }\bibfield  {title} {\bibinfo {title} {{The deep space quantum link: prospective fundamental physics experiments using long-baseline quantum optics}},\ }\href@noop {} {\bibfield  {journal} {\bibinfo  {journal} {EPJ Quantum Technol.}\ }\textbf {\bibinfo {volume} {9}} (\bibinfo {year} {2022})}\BibitemShut {NoStop}%
\bibitem [{\citenamefont {Horn}\ \emph {et~al.}(2012)\citenamefont {Horn} \emph {et~al.}}]{horn}%
  \BibitemOpen
  \bibfield  {author} {\bibinfo {author} {\bibfnamefont {R.}~\bibnamefont {Horn}} \emph {et~al.},\ }\bibfield  {title} {\bibinfo {title} {{Monolithic Source of Photon Pairs}},\ }\href@noop {} {\bibfield  {journal} {\bibinfo  {journal} {Phys. Rev. Lett.}\ }\textbf {\bibinfo {volume} {108}} (\bibinfo {year} {2012})}\BibitemShut {NoStop}%
\bibitem [{\citenamefont {Rudolph}(2017)}]{rudolph}%
  \BibitemOpen
  \bibfield  {author} {\bibinfo {author} {\bibfnamefont {T.}~\bibnamefont {Rudolph}},\ }\bibfield  {title} {\bibinfo {title} {{Why I am optimistic about the silicon-photonic route to quantum computing}},\ }\href@noop {} {\bibfield  {journal} {\bibinfo  {journal} {APL Photon.}\ }\textbf {\bibinfo {volume} {2}} (\bibinfo {year} {2017})}\BibitemShut {NoStop}%
\bibitem [{\citenamefont {Newman}\ \emph {et~al.}(2019)\citenamefont {Newman} \emph {et~al.}}]{newman}%
  \BibitemOpen
  \bibfield  {author} {\bibinfo {author} {\bibfnamefont {Z.}~\bibnamefont {Newman}} \emph {et~al.},\ }\bibfield  {title} {\bibinfo {title} {{Architecture for the photonic integration of an optical atomic clock}},\ }\href@noop {} {\bibfield  {journal} {\bibinfo  {journal} {Optica}\ }\textbf {\bibinfo {volume} {6}} (\bibinfo {year} {2019})}\BibitemShut {NoStop}%
\bibitem [{\citenamefont {Matsuda}\ \emph {et~al.}(2012)\citenamefont {Matsuda} \emph {et~al.}}]{matsuda}%
  \BibitemOpen
  \bibfield  {author} {\bibinfo {author} {\bibfnamefont {N.}~\bibnamefont {Matsuda}} \emph {et~al.},\ }\bibfield  {title} {\bibinfo {title} {{A monolithically integrated polarization entangled photon pair source on a silicon chip}},\ }\href@noop {} {\bibfield  {journal} {\bibinfo  {journal} {Sci. Rep.}\ }\textbf {\bibinfo {volume} {2}} (\bibinfo {year} {2012})}\BibitemShut {NoStop}%
\bibitem [{\citenamefont {{PsiQuantum Team}}(2025)}]{psi}%
  \BibitemOpen
  \bibfield  {author} {\bibinfo {author} {\bibnamefont {{PsiQuantum Team}}},\ }\bibfield  {title} {\bibinfo {title} {{A manufacturable platform for photonic quantum computing}},\ }\href@noop {} {\bibfield  {journal} {\bibinfo  {journal} {Nature}\ }\textbf {\bibinfo {volume} {641}} (\bibinfo {year} {2025})}\BibitemShut {NoStop}%
\bibitem [{\citenamefont {Silverstone}\ \emph {et~al.}(2016)\citenamefont {Silverstone} \emph {et~al.}}]{silverstone2}%
  \BibitemOpen
  \bibfield  {author} {\bibinfo {author} {\bibfnamefont {J.}~\bibnamefont {Silverstone}} \emph {et~al.},\ }\bibfield  {title} {\bibinfo {title} {{Silicon Quantum Photonics}},\ }\href@noop {} {\bibfield  {journal} {\bibinfo  {journal} {IEEE J. Sel. Top. Quantum Electron.}\ }\textbf {\bibinfo {volume} {22}} (\bibinfo {year} {2016})}\BibitemShut {NoStop}%
\bibitem [{\citenamefont {DiVincenzo}(2000)}]{divincenzo}%
  \BibitemOpen
  \bibfield  {author} {\bibinfo {author} {\bibfnamefont {D.}~\bibnamefont {DiVincenzo}},\ }\bibfield  {title} {\bibinfo {title} {{The Physical Implementation of Quantum Computation}},\ }\href@noop {} {\bibfield  {journal} {\bibinfo  {journal} {Fortschr. Phys.}\ }\textbf {\bibinfo {volume} {48}} (\bibinfo {year} {2000})}\BibitemShut {NoStop}%
\bibitem [{\citenamefont {Collins}\ \emph {et~al.}(2013)\citenamefont {Collins} \emph {et~al.}}]{collins}%
  \BibitemOpen
  \bibfield  {author} {\bibinfo {author} {\bibfnamefont {M.}~\bibnamefont {Collins}} \emph {et~al.},\ }\bibfield  {title} {\bibinfo {title} {{Integrated spatial multiplexing of heralded single-photon sources}},\ }\href@noop {} {\bibfield  {journal} {\bibinfo  {journal} {Nat. Commun.}\ }\textbf {\bibinfo {volume} {4}} (\bibinfo {year} {2013})}\BibitemShut {NoStop}%
\bibitem [{\citenamefont {Ding}\ \emph {et~al.}(2025)\citenamefont {Ding} \emph {et~al.}}]{ding}%
  \BibitemOpen
  \bibfield  {author} {\bibinfo {author} {\bibfnamefont {X.}~\bibnamefont {Ding}} \emph {et~al.},\ }\bibfield  {title} {\bibinfo {title} {{High-efficiency single-photon source above the loss-tolerant threshold for efficient linear optical quantum computing}},\ }\href@noop {} {\bibfield  {journal} {\bibinfo  {journal} {Nat. Photon.}\ }\textbf {\bibinfo {volume} {19}} (\bibinfo {year} {2025})}\BibitemShut {NoStop}%
\bibitem [{\citenamefont {Abudayyeh}\ \emph {et~al.}(2021)\citenamefont {Abudayyeh} \emph {et~al.}}]{abudayyeh}%
  \BibitemOpen
  \bibfield  {author} {\bibinfo {author} {\bibfnamefont {H.}~\bibnamefont {Abudayyeh}} \emph {et~al.},\ }\bibfield  {title} {\bibinfo {title} {{Single photon sources with near unity collection efficiencies by deterministic placement of quantum dots in nanoantennas}},\ }\href@noop {} {\bibfield  {journal} {\bibinfo  {journal} {APL Photon.}\ }\textbf {\bibinfo {volume} {6}} (\bibinfo {year} {2021})}\BibitemShut {NoStop}%
\bibitem [{\citenamefont {Kala}\ \emph {et~al.}(2020)\citenamefont {Kala} \emph {et~al.}}]{kala}%
  \BibitemOpen
  \bibfield  {author} {\bibinfo {author} {\bibfnamefont {A.}~\bibnamefont {Kala}} \emph {et~al.},\ }\bibfield  {title} {\bibinfo {title} {{Hyperbolic Metamaterial with Quantum Dots for Enhanced Emission and Collection Efficiencies}},\ }\href@noop {} {\bibfield  {journal} {\bibinfo  {journal} {Adv. Optical Mater.}\ }\textbf {\bibinfo {volume} {8}} (\bibinfo {year} {2020})}\BibitemShut {NoStop}%
\bibitem [{\citenamefont {Li}\ \emph {et~al.}(2023)\citenamefont {Li} \emph {et~al.}}]{li}%
  \BibitemOpen
  \bibfield  {author} {\bibinfo {author} {\bibfnamefont {S.}~\bibnamefont {Li}} \emph {et~al.},\ }\bibfield  {title} {\bibinfo {title} {{Scalable Deterministic Integration of Two Quantum Dots into an On-Chip Quantum Circuit}},\ }\href@noop {} {\bibfield  {journal} {\bibinfo  {journal} {ACS Photonics}\ }\textbf {\bibinfo {volume} {10}} (\bibinfo {year} {2023})}\BibitemShut {NoStop}%
\bibitem [{\citenamefont {Liang}\ \emph {et~al.}(2022)\citenamefont {Liang} \emph {et~al.}}]{zhai}%
  \BibitemOpen
  \bibfield  {author} {\bibinfo {author} {\bibfnamefont {Z.}~\bibnamefont {Liang}} \emph {et~al.},\ }\bibfield  {title} {\bibinfo {title} {{Quantum interference of identical photons from remote GaAs quantum dots}},\ }\href@noop {} {\bibfield  {journal} {\bibinfo  {journal} {Nat. Nanotechnol.}\ }\textbf {\bibinfo {volume} {17}} (\bibinfo {year} {2022})}\BibitemShut {NoStop}%
\bibitem [{\citenamefont {Vajner}\ \emph {et~al.}(2024)\citenamefont {Vajner} \emph {et~al.}}]{vajner}%
  \BibitemOpen
  \bibfield  {author} {\bibinfo {author} {\bibfnamefont {D.}~\bibnamefont {Vajner}} \emph {et~al.},\ }\bibfield  {title} {\bibinfo {title} {{On-Demand Generation of Indistinguishable Photons in the Telecom C‑Band Using Quantum Dot Devices}},\ }\href@noop {} {\bibfield  {journal} {\bibinfo  {journal} {ACS Photonics}\ }\textbf {\bibinfo {volume} {11}} (\bibinfo {year} {2024})}\BibitemShut {NoStop}%
\bibitem [{\citenamefont {Lio}\ \emph {et~al.}(2022)\citenamefont {Lio} \emph {et~al.}}]{lio}%
  \BibitemOpen
  \bibfield  {author} {\bibinfo {author} {\bibfnamefont {B.}~\bibnamefont {Lio}} \emph {et~al.},\ }\bibfield  {title} {\bibinfo {title} {{A Pure and Indistinguishable Single-Photon Source at Telecommunication Wavelength}},\ }\href@noop {} {\bibfield  {journal} {\bibinfo  {journal} {Adv. Quantum Technol.}\ }\textbf {\bibinfo {volume} {5}} (\bibinfo {year} {2022})}\BibitemShut {NoStop}%
\bibitem [{\citenamefont {Yu}\ \emph {et~al.}(2023)\citenamefont {Yu} \emph {et~al.}}]{yu}%
  \BibitemOpen
  \bibfield  {author} {\bibinfo {author} {\bibfnamefont {Y.}~\bibnamefont {Yu}} \emph {et~al.},\ }\bibfield  {title} {\bibinfo {title} {{Telecom-band quantum dot technologies for long-distance quantum networks}},\ }\href@noop {} {\bibfield  {journal} {\bibinfo  {journal} {Nat. Nanotechnol.}\ }\textbf {\bibinfo {volume} {18}} (\bibinfo {year} {2023})}\BibitemShut {NoStop}%
\bibitem [{\citenamefont {Hepp}\ \emph {et~al.}(2019)\citenamefont {Hepp} \emph {et~al.}}]{hepp}%
  \BibitemOpen
  \bibfield  {author} {\bibinfo {author} {\bibfnamefont {S.}~\bibnamefont {Hepp}} \emph {et~al.},\ }\bibfield  {title} {\bibinfo {title} {{Semiconductor Quantum Dots for Integrated Quantum Photonics}},\ }\href@noop {} {\bibfield  {journal} {\bibinfo  {journal} {Adv. Quantum Technol.}\ }\textbf {\bibinfo {volume} {2}} (\bibinfo {year} {2019})}\BibitemShut {NoStop}%
\bibitem [{\citenamefont {Faruque}\ \emph {et~al.}(2018{\natexlab{a}})\citenamefont {Faruque} \emph {et~al.}}]{faruque2}%
  \BibitemOpen
  \bibfield  {author} {\bibinfo {author} {\bibfnamefont {I.}~\bibnamefont {Faruque}} \emph {et~al.},\ }\bibfield  {title} {\bibinfo {title} {{On-chip quantum interference with heralded photons from two independent micro-ring resonator sources in silicon photonics}},\ }\href@noop {} {\bibfield  {journal} {\bibinfo  {journal} {Opt. Express}\ }\textbf {\bibinfo {volume} {26}} (\bibinfo {year} {2018}{\natexlab{a}})}\BibitemShut {NoStop}%
\bibitem [{\citenamefont {Faruque}\ \emph {et~al.}(2019)\citenamefont {Faruque} \emph {et~al.}}]{faruque3}%
  \BibitemOpen
  \bibfield  {author} {\bibinfo {author} {\bibfnamefont {I.}~\bibnamefont {Faruque}} \emph {et~al.},\ }\bibfield  {title} {\bibinfo {title} {{Estimating the Indistinguishability of Heralded Single Photons Using Second-Order Correlation}},\ }\href@noop {} {\bibfield  {journal} {\bibinfo  {journal} {Phys. Rev. Appl.}\ }\textbf {\bibinfo {volume} {12}} (\bibinfo {year} {2019})}\BibitemShut {NoStop}%
\bibitem [{\citenamefont {Spring}\ \emph {et~al.}(2017)\citenamefont {Spring} \emph {et~al.}}]{spring}%
  \BibitemOpen
  \bibfield  {author} {\bibinfo {author} {\bibfnamefont {J.}~\bibnamefont {Spring}} \emph {et~al.},\ }\bibfield  {title} {\bibinfo {title} {{Chip-based array of near-identical, pure, heralded single-photon sources}},\ }\href@noop {} {\bibfield  {journal} {\bibinfo  {journal} {Optica}\ }\textbf {\bibinfo {volume} {4}} (\bibinfo {year} {2017})}\BibitemShut {NoStop}%
\bibitem [{\citenamefont {Llewellyn}\ \emph {et~al.}(2020)\citenamefont {Llewellyn} \emph {et~al.}}]{llewellyn}%
  \BibitemOpen
  \bibfield  {author} {\bibinfo {author} {\bibfnamefont {D.}~\bibnamefont {Llewellyn}} \emph {et~al.},\ }\bibfield  {title} {\bibinfo {title} {{Chip-to-chip quantum teleportation and multi-photon entanglement in silicon}},\ }\href@noop {} {\bibfield  {journal} {\bibinfo  {journal} {Nat. Phys.}\ }\textbf {\bibinfo {volume} {16}} (\bibinfo {year} {2020})}\BibitemShut {NoStop}%
\bibitem [{\citenamefont {Zhu}\ \emph {et~al.}(2020)\citenamefont {Zhu} \emph {et~al.}}]{zhu}%
  \BibitemOpen
  \bibfield  {author} {\bibinfo {author} {\bibfnamefont {D.}~\bibnamefont {Zhu}} \emph {et~al.},\ }\bibfield  {title} {\bibinfo {title} {{Resolving Photon Numbers Using a Superconducting Nanowire with Impedance-Matching Taper}},\ }\href@noop {} {\bibfield  {journal} {\bibinfo  {journal} {Nano Lett.}\ }\textbf {\bibinfo {volume} {20}} (\bibinfo {year} {2020})}\BibitemShut {NoStop}%
\bibitem [{\citenamefont {Vernon}\ \emph {et~al.}(2017)\citenamefont {Vernon} \emph {et~al.}}]{vernon}%
  \BibitemOpen
  \bibfield  {author} {\bibinfo {author} {\bibfnamefont {Z.}~\bibnamefont {Vernon}} \emph {et~al.},\ }\bibfield  {title} {\bibinfo {title} {{Truly unentangled photon pairs without spectral filtering}},\ }\href@noop {} {\bibfield  {journal} {\bibinfo  {journal} {Opt. Lett.}\ }\textbf {\bibinfo {volume} {42}} (\bibinfo {year} {2017})}\BibitemShut {NoStop}%
\bibitem [{\citenamefont {Burridge}\ \emph {et~al.}(2020)\citenamefont {Burridge} \emph {et~al.}}]{ben1}%
  \BibitemOpen
  \bibfield  {author} {\bibinfo {author} {\bibfnamefont {B.}~\bibnamefont {Burridge}} \emph {et~al.},\ }\bibfield  {title} {\bibinfo {title} {{High spectro-temporal purity single-photons from silicon micro-racetrack resonators using a dual-pulse configuration}},\ }\href@noop {} {\bibfield  {journal} {\bibinfo  {journal} {Opt. Lett.}\ }\textbf {\bibinfo {volume} {45}} (\bibinfo {year} {2020})}\BibitemShut {NoStop}%
\bibitem [{\citenamefont {Christensen}\ \emph {et~al.}(2018)\citenamefont {Christensen} \emph {et~al.}}]{christensen}%
  \BibitemOpen
  \bibfield  {author} {\bibinfo {author} {\bibfnamefont {J.}~\bibnamefont {Christensen}} \emph {et~al.},\ }\bibfield  {title} {\bibinfo {title} {{Engineering spectrally unentangled photon pairs from nonlinear microring resonators by pump manipulation}},\ }\href@noop {} {\bibfield  {journal} {\bibinfo  {journal} {Opt. Lett.}\ }\textbf {\bibinfo {volume} {43}} (\bibinfo {year} {2018})}\BibitemShut {NoStop}%
\bibitem [{\citenamefont {Yin}\ \emph {et~al.}(2020)\citenamefont {Yin} \emph {et~al.}}]{yin}%
  \BibitemOpen
  \bibfield  {author} {\bibinfo {author} {\bibfnamefont {J.}~\bibnamefont {Yin}} \emph {et~al.},\ }\bibfield  {title} {\bibinfo {title} {{Entanglement-based secure quantum cryptography over 1,120 kilometres}},\ }\href@noop {} {\bibfield  {journal} {\bibinfo  {journal} {Nature}\ }\textbf {\bibinfo {volume} {582}} (\bibinfo {year} {2020})}\BibitemShut {NoStop}%
\bibitem [{\citenamefont {Sparrow}\ \emph {et~al.}(2018)\citenamefont {Sparrow} \emph {et~al.}}]{sparrow}%
  \BibitemOpen
  \bibfield  {author} {\bibinfo {author} {\bibfnamefont {C.}~\bibnamefont {Sparrow}} \emph {et~al.},\ }\bibfield  {title} {\bibinfo {title} {{Simulating the vibrational quantum dynamics of molecules using photonics}},\ }\href@noop {} {\bibfield  {journal} {\bibinfo  {journal} {Nature}\ }\textbf {\bibinfo {volume} {557}} (\bibinfo {year} {2018})}\BibitemShut {NoStop}%
\bibitem [{\citenamefont {Spedalieri}\ \emph {et~al.}(2020)\citenamefont {Spedalieri} \emph {et~al.}}]{spedalieri}%
  \BibitemOpen
  \bibfield  {author} {\bibinfo {author} {\bibfnamefont {G.}~\bibnamefont {Spedalieri}} \emph {et~al.},\ }\bibfield  {title} {\bibinfo {title} {{Detecting and tracking bacteria with quantum light}},\ }\href@noop {} {\bibfield  {journal} {\bibinfo  {journal} {Phys. Rev. Res.}\ }\textbf {\bibinfo {volume} {2}} (\bibinfo {year} {2020})}\BibitemShut {NoStop}%
\bibitem [{\citenamefont {Daher}\ \emph {et~al.}(2021)\citenamefont {Daher} \emph {et~al.}}]{daher}%
  \BibitemOpen
  \bibfield  {author} {\bibinfo {author} {\bibfnamefont {M.}~\bibnamefont {Daher}} \emph {et~al.},\ }\bibfield  {title} {\bibinfo {title} {{Design of a novel optical sensor for the detection of waterborne bacteria based on a photonic crystal with an ultra‑high sensitivity}},\ }\href@noop {} {\bibfield  {journal} {\bibinfo  {journal} {Opt. Quant. Electron.}\ }\textbf {\bibinfo {volume} {54}} (\bibinfo {year} {2021})}\BibitemShut {NoStop}%
\bibitem [{\citenamefont {Bleuler}\ and\ \citenamefont {Bradt}(1948)}]{bleuler}%
  \BibitemOpen
  \bibfield  {author} {\bibinfo {author} {\bibfnamefont {E.}~\bibnamefont {Bleuler}}\ and\ \bibinfo {author} {\bibfnamefont {H.}~\bibnamefont {Bradt}},\ }\bibfield  {title} {\bibinfo {title} {{Correlation between the States of Polarization of the Two Quanta of Annihilation Radiation}},\ }\href@noop {} {\bibfield  {journal} {\bibinfo  {journal} {Phys. Rev.}\ }\textbf {\bibinfo {volume} {73}} (\bibinfo {year} {1948})}\BibitemShut {NoStop}%
\bibitem [{\citenamefont {Burnham}\ and\ \citenamefont {Weinberg}(1970)}]{burnham}%
  \BibitemOpen
  \bibfield  {author} {\bibinfo {author} {\bibfnamefont {D.}~\bibnamefont {Burnham}}\ and\ \bibinfo {author} {\bibfnamefont {D.}~\bibnamefont {Weinberg}},\ }\bibfield  {title} {\bibinfo {title} {{Observation of simultaneity in parametric production of optical photon pairs}},\ }\href@noop {} {\bibfield  {journal} {\bibinfo  {journal} {Phys. Rev. Lett.}\ }\textbf {\bibinfo {volume} {25}} (\bibinfo {year} {1970})}\BibitemShut {NoStop}%
\bibitem [{\citenamefont {Bonfrate}\ \emph {et~al.}(1999)\citenamefont {Bonfrate} \emph {et~al.}}]{bonfrate}%
  \BibitemOpen
  \bibfield  {author} {\bibinfo {author} {\bibfnamefont {G.}~\bibnamefont {Bonfrate}} \emph {et~al.},\ }\bibfield  {title} {\bibinfo {title} {{Parametric fluorescence in periodically poled silica fibers}},\ }\href@noop {} {\bibfield  {journal} {\bibinfo  {journal} {Appl. Phys. Lett.}\ }\textbf {\bibinfo {volume} {75}} (\bibinfo {year} {1999})}\BibitemShut {NoStop}%
\bibitem [{\citenamefont {Fukuda}\ \emph {et~al.}(2005)\citenamefont {Fukuda} \emph {et~al.}}]{fukuda}%
  \BibitemOpen
  \bibfield  {author} {\bibinfo {author} {\bibfnamefont {H.}~\bibnamefont {Fukuda}} \emph {et~al.},\ }\bibfield  {title} {\bibinfo {title} {{Four-wave mixing in silicon wire waveguides}},\ }\href@noop {} {\bibfield  {journal} {\bibinfo  {journal} {Opt. Express}\ }\textbf {\bibinfo {volume} {13}} (\bibinfo {year} {2005})}\BibitemShut {NoStop}%
\bibitem [{\citenamefont {Sharping}\ \emph {et~al.}(2004)\citenamefont {Sharping} \emph {et~al.}}]{sharping}%
  \BibitemOpen
  \bibfield  {author} {\bibinfo {author} {\bibfnamefont {J.}~\bibnamefont {Sharping}} \emph {et~al.},\ }\bibfield  {title} {\bibinfo {title} {{Quantum-correlated twin photons from microstructure fiber}},\ }\href@noop {} {\bibfield  {journal} {\bibinfo  {journal} {Opt. Express}\ }\textbf {\bibinfo {volume} {12}} (\bibinfo {year} {2004})}\BibitemShut {NoStop}%
\bibitem [{\citenamefont {Suzuki}\ \emph {et~al.}(2009)\citenamefont {Suzuki}, \citenamefont {Hamachi},\ and\ \citenamefont {Baba}}]{suzuki}%
  \BibitemOpen
  \bibfield  {author} {\bibinfo {author} {\bibfnamefont {K.}~\bibnamefont {Suzuki}}, \bibinfo {author} {\bibfnamefont {Y.}~\bibnamefont {Hamachi}},\ and\ \bibinfo {author} {\bibfnamefont {T.}~\bibnamefont {Baba}},\ }\bibfield  {title} {\bibinfo {title} {{Fabrication and characterization of chalcogenide glass photonic crystal waveguides}},\ }\href@noop {} {\bibfield  {journal} {\bibinfo  {journal} {Opt. Express}\ }\textbf {\bibinfo {volume} {17}} (\bibinfo {year} {2009})}\BibitemShut {NoStop}%
\bibitem [{\citenamefont {Clemmen}\ \emph {et~al.}(2009)\citenamefont {Clemmen} \emph {et~al.}}]{clemmen}%
  \BibitemOpen
  \bibfield  {author} {\bibinfo {author} {\bibfnamefont {S.}~\bibnamefont {Clemmen}} \emph {et~al.},\ }\bibfield  {title} {\bibinfo {title} {{Continuous wave photon pair generation in silicon-on-insulator waveguides and ring resonators}},\ }\href@noop {} {\bibfield  {journal} {\bibinfo  {journal} {Opt. Express}\ }\textbf {\bibinfo {volume} {17}} (\bibinfo {year} {2009})}\BibitemShut {NoStop}%
\bibitem [{\citenamefont {Zeng}\ \emph {et~al.}(2014)\citenamefont {Zeng}, \citenamefont {Gentry},\ and\ \citenamefont {Popović}}]{zeng}%
  \BibitemOpen
  \bibfield  {author} {\bibinfo {author} {\bibfnamefont {X.}~\bibnamefont {Zeng}}, \bibinfo {author} {\bibfnamefont {C.}~\bibnamefont {Gentry}},\ and\ \bibinfo {author} {\bibfnamefont {M.}~\bibnamefont {Popović}},\ }\bibfield  {title} {\bibinfo {title} {{Four-wave mixing in silicon “photonic molecule” resonators with port-selective, orthogonal supermode excitation}},\ }in\ \href@noop {} {\emph {\bibinfo {booktitle} {Conference on Lasers and Electro-Optics (CLEO), OSA Technical Digest}}}\ (\bibinfo {year} {2014})\BibitemShut {NoStop}%
\bibitem [{\citenamefont {Liu}\ \emph {et~al.}(2020)\citenamefont {Liu} \emph {et~al.}}]{liu}%
  \BibitemOpen
  \bibfield  {author} {\bibinfo {author} {\bibfnamefont {Y.}~\bibnamefont {Liu}} \emph {et~al.},\ }\bibfield  {title} {\bibinfo {title} {{High-spectral-purity photon generation from a dual-interferometer-coupled silicon microring}},\ }\href@noop {} {\bibfield  {journal} {\bibinfo  {journal} {Opt. Lett.}\ }\textbf {\bibinfo {volume} {45}} (\bibinfo {year} {2020})}\BibitemShut {NoStop}%
\bibitem [{\citenamefont {Burridge}\ \emph {et~al.}(2023)\citenamefont {Burridge} \emph {et~al.}}]{integrateandscale}%
  \BibitemOpen
  \bibfield  {author} {\bibinfo {author} {\bibfnamefont {B.}~\bibnamefont {Burridge}} \emph {et~al.},\ }\bibfield  {title} {\bibinfo {title} {{Integrate and scale: a source of spectrally separable photon pairs}},\ }\href@noop {} {\bibfield  {journal} {\bibinfo  {journal} {Optica}\ }\textbf {\bibinfo {volume} {10}} (\bibinfo {year} {2023})}\BibitemShut {NoStop}%
\bibitem [{\citenamefont {Kaneda}\ and\ \citenamefont {Kwiat}(2019)}]{kaneda}%
  \BibitemOpen
  \bibfield  {author} {\bibinfo {author} {\bibfnamefont {F.}~\bibnamefont {Kaneda}}\ and\ \bibinfo {author} {\bibfnamefont {P.}~\bibnamefont {Kwiat}},\ }\bibfield  {title} {\bibinfo {title} {{High-efficiency single-photon generation via large-scale active time multiplexing}},\ }\href@noop {} {\bibfield  {journal} {\bibinfo  {journal} {Sci. Adv.}\ }\textbf {\bibinfo {volume} {5}} (\bibinfo {year} {2019})}\BibitemShut {NoStop}%
\bibitem [{\citenamefont {Meyer-Scott}\ \emph {et~al.}(2020)\citenamefont {Meyer-Scott} \emph {et~al.}}]{meyer-scott}%
  \BibitemOpen
  \bibfield  {author} {\bibinfo {author} {\bibfnamefont {E.}~\bibnamefont {Meyer-Scott}} \emph {et~al.},\ }\bibfield  {title} {\bibinfo {title} {{Single-photon sources: Approaching the ideal through multiplexing}},\ }\href@noop {} {\bibfield  {journal} {\bibinfo  {journal} {Rev. Sci. Instrum.}\ }\textbf {\bibinfo {volume} {91}} (\bibinfo {year} {2020})}\BibitemShut {NoStop}%
\bibitem [{\citenamefont {Kaneda}\ \emph {et~al.}(2016)\citenamefont {Kaneda} \emph {et~al.}}]{kaneda2}%
  \BibitemOpen
  \bibfield  {author} {\bibinfo {author} {\bibfnamefont {F.}~\bibnamefont {Kaneda}} \emph {et~al.},\ }\bibfield  {title} {\bibinfo {title} {{Heralded single-photon source utilizing highly nondegenerate, spectrally factorable spontaneous parametric downconversion}},\ }\href@noop {} {\bibfield  {journal} {\bibinfo  {journal} {Opt. Express}\ }\textbf {\bibinfo {volume} {24}} (\bibinfo {year} {2016})}\BibitemShut {NoStop}%
\bibitem [{\citenamefont {Cao}\ \emph {et~al.}(2018)\citenamefont {Cao} \emph {et~al.}}]{cao}%
  \BibitemOpen
  \bibfield  {author} {\bibinfo {author} {\bibfnamefont {Y.}~\bibnamefont {Cao}} \emph {et~al.},\ }\bibfield  {title} {\bibinfo {title} {{Bell Test over Extremely High-Loss Channels: Towards Distributing Entangled Photon Pairs between Earth and the Moon}},\ }\href@noop {} {\bibfield  {journal} {\bibinfo  {journal} {Phys. Rev. Lett.}\ }\textbf {\bibinfo {volume} {120}} (\bibinfo {year} {2018})}\BibitemShut {NoStop}%
\bibitem [{\citenamefont {Liu}\ and\ \citenamefont {Lim}(2014)}]{liulim}%
  \BibitemOpen
  \bibfield  {author} {\bibinfo {author} {\bibfnamefont {M.}~\bibnamefont {Liu}}\ and\ \bibinfo {author} {\bibfnamefont {H.}~\bibnamefont {Lim}},\ }\bibfield  {title} {\bibinfo {title} {{Efficient heralding of O-band passively spatial-multiplexed photons for noise-tolerant quantum key distribution}},\ }\href@noop {} {\bibfield  {journal} {\bibinfo  {journal} {Opt. Express}\ }\textbf {\bibinfo {volume} {22}} (\bibinfo {year} {2014})}\BibitemShut {NoStop}%
\bibitem [{\citenamefont {Meyer-Scott}\ \emph {et~al.}(2017)\citenamefont {Meyer-Scott} \emph {et~al.}}]{meyer-scott2}%
  \BibitemOpen
  \bibfield  {author} {\bibinfo {author} {\bibfnamefont {E.}~\bibnamefont {Meyer-Scott}} \emph {et~al.},\ }\bibfield  {title} {\bibinfo {title} {{Limits on the heralding efficiencies and spectral purities of spectrally filtered single photons from photon-pair sources}},\ }\href@noop {} {\bibfield  {journal} {\bibinfo  {journal} {Phys. Rev. A}\ }\textbf {\bibinfo {volume} {95}} (\bibinfo {year} {2017})}\BibitemShut {NoStop}%
\bibitem [{\citenamefont {Neville}\ \emph {et~al.}(2017)\citenamefont {Neville} \emph {et~al.}}]{neville}%
  \BibitemOpen
  \bibfield  {author} {\bibinfo {author} {\bibfnamefont {A.}~\bibnamefont {Neville}} \emph {et~al.},\ }\bibfield  {title} {\bibinfo {title} {{Classical boson sampling algorithms with superior performance to near-term experiments}},\ }\href@noop {} {\bibfield  {journal} {\bibinfo  {journal} {Nat. Phys.}\ }\textbf {\bibinfo {volume} {13}} (\bibinfo {year} {2017})}\BibitemShut {NoStop}%
\bibitem [{\citenamefont {Zhong}\ \emph {et~al.}(2020)\citenamefont {Zhong} \emph {et~al.}}]{zhong}%
  \BibitemOpen
  \bibfield  {author} {\bibinfo {author} {\bibfnamefont {H.}~\bibnamefont {Zhong}} \emph {et~al.},\ }\bibfield  {title} {\bibinfo {title} {{Quantum computational advantage using photons}},\ }\href@noop {} {\bibfield  {journal} {\bibinfo  {journal} {Science}\ }\textbf {\bibinfo {volume} {370}} (\bibinfo {year} {2020})}\BibitemShut {NoStop}%
\bibitem [{\citenamefont {Steinlechner}\ \emph {et~al.}(2012)\citenamefont {Steinlechner} \emph {et~al.}}]{steinlechner}%
  \BibitemOpen
  \bibfield  {author} {\bibinfo {author} {\bibfnamefont {F.}~\bibnamefont {Steinlechner}} \emph {et~al.},\ }\bibfield  {title} {\bibinfo {title} {{A high-brightness source of polarization-entangled photons optimized for applications in free space}},\ }\href@noop {} {\bibfield  {journal} {\bibinfo  {journal} {Opt. Express}\ }\textbf {\bibinfo {volume} {20}} (\bibinfo {year} {2012})}\BibitemShut {NoStop}%
\bibitem [{\citenamefont {Ling}\ \emph {et~al.}(2024)\citenamefont {Ling} \emph {et~al.}}]{ling}%
  \BibitemOpen
  \bibfield  {author} {\bibinfo {author} {\bibfnamefont {A.}~\bibnamefont {Ling}} \emph {et~al.},\ }\bibfield  {title} {\bibinfo {title} {{Demonstration of entanglement distribution over 155 km metropolitan fiber using a silicon nanophotonic chip}},\ }\href@noop {} {\bibfield  {journal} {\bibinfo  {journal} {arXiv:2409.17558 [quant-ph]}\ } (\bibinfo {year} {2024})}\BibitemShut {NoStop}%
\bibitem [{\citenamefont {Helt}\ \emph {et~al.}(2010)\citenamefont {Helt} \emph {et~al.}}]{helt}%
  \BibitemOpen
  \bibfield  {author} {\bibinfo {author} {\bibfnamefont {L.}~\bibnamefont {Helt}} \emph {et~al.},\ }\bibfield  {title} {\bibinfo {title} {{Spontaneous four-wave mixing in microring resonators}},\ }\href@noop {} {\bibfield  {journal} {\bibinfo  {journal} {Opt. Lett.}\ }\textbf {\bibinfo {volume} {35}} (\bibinfo {year} {2010})}\BibitemShut {NoStop}%
\bibitem [{\citenamefont {Husko}\ \emph {et~al.}(2013)\citenamefont {Husko} \emph {et~al.}}]{husko}%
  \BibitemOpen
  \bibfield  {author} {\bibinfo {author} {\bibfnamefont {C.}~\bibnamefont {Husko}} \emph {et~al.},\ }\bibfield  {title} {\bibinfo {title} {{Multi-photon absorption limits to heralded single photon sources}},\ }\href@noop {} {\bibfield  {journal} {\bibinfo  {journal} {Sci. Rep.}\ }\textbf {\bibinfo {volume} {3}} (\bibinfo {year} {2013})}\BibitemShut {NoStop}%
\bibitem [{\citenamefont {Engin}\ \emph {et~al.}(2013)\citenamefont {Engin} \emph {et~al.}}]{engin}%
  \BibitemOpen
  \bibfield  {author} {\bibinfo {author} {\bibfnamefont {E.}~\bibnamefont {Engin}} \emph {et~al.},\ }\bibfield  {title} {\bibinfo {title} {{Photon pair generation in a silicon micro-ring resonator with reverse bias enhancement}},\ }\href@noop {} {\bibfield  {journal} {\bibinfo  {journal} {Opt. Express}\ }\textbf {\bibinfo {volume} {21}} (\bibinfo {year} {2013})}\BibitemShut {NoStop}%
\bibitem [{\citenamefont {He}\ \emph {et~al.}(2020)\citenamefont {He} \emph {et~al.}}]{couplers}%
  \BibitemOpen
  \bibfield  {author} {\bibinfo {author} {\bibfnamefont {A.}~\bibnamefont {He}} \emph {et~al.},\ }\bibfield  {title} {\bibinfo {title} {{Low Loss, Large Bandwidth Fiber-Chip Edge Couplers Based on Silicon-on-Insulator Platform}},\ }\href@noop {} {\bibfield  {journal} {\bibinfo  {journal} {J. Lightwave Technol.}\ }\textbf {\bibinfo {volume} {38}} (\bibinfo {year} {2020})}\BibitemShut {NoStop}%
\bibitem [{\citenamefont {Du}\ \emph {et~al.}(2024)\citenamefont {Du} \emph {et~al.}}]{du}%
  \BibitemOpen
  \bibfield  {author} {\bibinfo {author} {\bibfnamefont {J.}~\bibnamefont {Du}} \emph {et~al.},\ }\bibfield  {title} {\bibinfo {title} {{Demonstration of a low loss, highly stable and re-useable edge coupler for high heralding efficiency and low g$^{(2)}$(0) SOI correlated photon pair sources}},\ }\href@noop {} {\bibfield  {journal} {\bibinfo  {journal} {Opt. Express}\ }\textbf {\bibinfo {volume} {32}} (\bibinfo {year} {2024})}\BibitemShut {NoStop}%
\bibitem [{\citenamefont {Paesani}\ \emph {et~al.}(2020)\citenamefont {Paesani} \emph {et~al.}}]{paesani}%
  \BibitemOpen
  \bibfield  {author} {\bibinfo {author} {\bibfnamefont {S.}~\bibnamefont {Paesani}} \emph {et~al.},\ }\bibfield  {title} {\bibinfo {title} {{Near-ideal spontaneous photon sources in silicon quantum photonics}},\ }\href@noop {} {\bibfield  {journal} {\bibinfo  {journal} {Nat. Commun.}\ }\textbf {\bibinfo {volume} {11}} (\bibinfo {year} {2020})}\BibitemShut {NoStop}%
\bibitem [{\citenamefont {Thomas}\ \emph {et~al.}(2021)\citenamefont {Thomas}, \citenamefont {McCutcheon},\ and\ \citenamefont {McCutcheon}}]{thomas}%
  \BibitemOpen
  \bibfield  {author} {\bibinfo {author} {\bibfnamefont {O.}~\bibnamefont {Thomas}}, \bibinfo {author} {\bibfnamefont {W.}~\bibnamefont {McCutcheon}},\ and\ \bibinfo {author} {\bibfnamefont {D.}~\bibnamefont {McCutcheon}},\ }\bibfield  {title} {\bibinfo {title} {{A general framework for multimode Gaussian quantum optics and photo-detection: Application to Hong–Ou–Mandel interference with filtered heralded single photon sources}},\ }\href@noop {} {\bibfield  {journal} {\bibinfo  {journal} {APL Photon.}\ }\textbf {\bibinfo {volume} {6}} (\bibinfo {year} {2021})}\BibitemShut {NoStop}%
\bibitem [{\citenamefont {Mičuda}\ \emph {et~al.}(2014)\citenamefont {Mičuda} \emph {et~al.}}]{micuda}%
  \BibitemOpen
  \bibfield  {author} {\bibinfo {author} {\bibfnamefont {M.}~\bibnamefont {Mičuda}} \emph {et~al.},\ }\bibfield  {title} {\bibinfo {title} {{Process-fidelity estimation of a linear optical quantum-controlled-Z gate: A comparative study}},\ }\href@noop {} {\bibfield  {journal} {\bibinfo  {journal} {Phys. Rev. A}\ }\textbf {\bibinfo {volume} {89}} (\bibinfo {year} {2014})}\BibitemShut {NoStop}%
\bibitem [{\citenamefont {Wallman}\ \emph {et~al.}(2015)\citenamefont {Wallman} \emph {et~al.}}]{wallman}%
  \BibitemOpen
  \bibfield  {author} {\bibinfo {author} {\bibfnamefont {J.}~\bibnamefont {Wallman}} \emph {et~al.},\ }\bibfield  {title} {\bibinfo {title} {{Estimating the coherence of noise}},\ }\href@noop {} {\bibfield  {journal} {\bibinfo  {journal} {New J. Phys.}\ }\textbf {\bibinfo {volume} {17}} (\bibinfo {year} {2015})}\BibitemShut {NoStop}%
\bibitem [{\citenamefont {Gaebler}\ \emph {et~al.}(2012)\citenamefont {Gaebler} \emph {et~al.}}]{gaebler}%
  \BibitemOpen
  \bibfield  {author} {\bibinfo {author} {\bibfnamefont {J.}~\bibnamefont {Gaebler}} \emph {et~al.},\ }\bibfield  {title} {\bibinfo {title} {{Randomized Benchmarking of Multiqubit Gates}},\ }\href@noop {} {\bibfield  {journal} {\bibinfo  {journal} {Phys. Rev. Lett.}\ }\textbf {\bibinfo {volume} {109}} (\bibinfo {year} {2012})}\BibitemShut {NoStop}%
\bibitem [{\citenamefont {Vernon}\ \emph {et~al.}(2016)\citenamefont {Vernon}, \citenamefont {Liscidini},\ and\ \citenamefont {Sipe}}]{lunch}%
  \BibitemOpen
  \bibfield  {author} {\bibinfo {author} {\bibfnamefont {Z.}~\bibnamefont {Vernon}}, \bibinfo {author} {\bibfnamefont {M.}~\bibnamefont {Liscidini}},\ and\ \bibinfo {author} {\bibfnamefont {J.}~\bibnamefont {Sipe}},\ }\bibfield  {title} {\bibinfo {title} {{No free lunch: the trade-off between heralding rate and efficiency in microresonator-based heralded single photon sources}},\ }\href@noop {} {\bibfield  {journal} {\bibinfo  {journal} {Opt. Lett.}\ }\textbf {\bibinfo {volume} {41}} (\bibinfo {year} {2016})}\BibitemShut {NoStop}%
\bibitem [{\citenamefont {Rodda}\ \emph {et~al.}(2024)\citenamefont {Rodda} \emph {et~al.}}]{rodda}%
  \BibitemOpen
  \bibfield  {author} {\bibinfo {author} {\bibfnamefont {L.}~\bibnamefont {Rodda}} \emph {et~al.},\ }\bibfield  {title} {\bibinfo {title} {{Effect on the spectral purity of photon-pairs due to on-chip temporal manipulation of the pump}},\ }\href@noop {} {\bibfield  {journal} {\bibinfo  {journal} {APL Quantum}\ }\textbf {\bibinfo {volume} {1}} (\bibinfo {year} {2024})}\BibitemShut {NoStop}%
\bibitem [{\citenamefont {Zielnicki}\ \emph {et~al.}(2018)\citenamefont {Zielnicki} \emph {et~al.}}]{zielnicki}%
  \BibitemOpen
  \bibfield  {author} {\bibinfo {author} {\bibfnamefont {K.}~\bibnamefont {Zielnicki}} \emph {et~al.},\ }\bibfield  {title} {\bibinfo {title} {{Joint spectral characterization of photon-pair sources}},\ }\href@noop {} {\bibfield  {journal} {\bibinfo  {journal} {J. Mod. Opt.}\ }\textbf {\bibinfo {volume} {65}} (\bibinfo {year} {2018})}\BibitemShut {NoStop}%
\bibitem [{\citenamefont {Moody}\ \emph {et~al.}(2022)\citenamefont {Moody} \emph {et~al.}}]{moody}%
  \BibitemOpen
  \bibfield  {author} {\bibinfo {author} {\bibfnamefont {G.}~\bibnamefont {Moody}} \emph {et~al.},\ }\bibfield  {title} {\bibinfo {title} {{2022 Roadmap on integrated quantum photonics}},\ }\href@noop {} {\bibfield  {journal} {\bibinfo  {journal} {J. Phys. Photonics}\ }\textbf {\bibinfo {volume} {4}} (\bibinfo {year} {2022})}\BibitemShut {NoStop}%
\bibitem [{\citenamefont {Ma}\ \emph {et~al.}(2017)\citenamefont {Ma} \emph {et~al.}}]{ma}%
  \BibitemOpen
  \bibfield  {author} {\bibinfo {author} {\bibfnamefont {C.}~\bibnamefont {Ma}} \emph {et~al.},\ }\bibfield  {title} {\bibinfo {title} {{Silicon photonic entangled photon-pair and heralded single photon generation with CAR \textgreater 12,000 and g$^{(2)}$(0) \textless 0.006}},\ }\href@noop {} {\bibfield  {journal} {\bibinfo  {journal} {Opt. Express}\ }\textbf {\bibinfo {volume} {25}} (\bibinfo {year} {2017})}\BibitemShut {NoStop}%
\bibitem [{\citenamefont {Esmann}\ \emph {et~al.}(2024)\citenamefont {Esmann}, \citenamefont {Wein},\ and\ \citenamefont {Antón-Solanas}}]{esmann}%
  \BibitemOpen
  \bibfield  {author} {\bibinfo {author} {\bibfnamefont {M.}~\bibnamefont {Esmann}}, \bibinfo {author} {\bibfnamefont {S.}~\bibnamefont {Wein}},\ and\ \bibinfo {author} {\bibfnamefont {C.}~\bibnamefont {Antón-Solanas}},\ }\bibfield  {title} {\bibinfo {title} {{Solid-State Single-Photon Sources: Recent Advances for Novel Quantum Materials}},\ }\href@noop {} {\bibfield  {journal} {\bibinfo  {journal} {Adv. Funct. Mater.}\ }\textbf {\bibinfo {volume} {34}} (\bibinfo {year} {2024})}\BibitemShut {NoStop}%
\bibitem [{\citenamefont {Wu}\ \emph {et~al.}(2022)\citenamefont {Wu} \emph {et~al.}}]{wu}%
  \BibitemOpen
  \bibfield  {author} {\bibinfo {author} {\bibfnamefont {C.}~\bibnamefont {Wu}} \emph {et~al.},\ }\bibfield  {title} {\bibinfo {title} {{Optimization of quantum light sources and four-wave mixing based on a reconfigurable silicon ring resonator}},\ }\href@noop {} {\bibfield  {journal} {\bibinfo  {journal} {Opt. Express}\ }\textbf {\bibinfo {volume} {30}} (\bibinfo {year} {2022})}\BibitemShut {NoStop}%
\bibitem [{\citenamefont {Steiner}\ \emph {et~al.}(2021)\citenamefont {Steiner} \emph {et~al.}}]{steiner}%
  \BibitemOpen
  \bibfield  {author} {\bibinfo {author} {\bibfnamefont {T.}~\bibnamefont {Steiner}} \emph {et~al.},\ }\bibfield  {title} {\bibinfo {title} {{Ultrabright Entangled-Photon-Pair Generation from an AlGaAs-On-Insulator Microring Resonator}},\ }\href@noop {} {\bibfield  {journal} {\bibinfo  {journal} {PRX Quantum}\ }\textbf {\bibinfo {volume} {2}} (\bibinfo {year} {2021})}\BibitemShut {NoStop}%
\bibitem [{\citenamefont {Silverstone}\ \emph {et~al.}(2015)\citenamefont {Silverstone} \emph {et~al.}}]{silverstone1}%
  \BibitemOpen
  \bibfield  {author} {\bibinfo {author} {\bibfnamefont {J.}~\bibnamefont {Silverstone}} \emph {et~al.},\ }\bibfield  {title} {\bibinfo {title} {{Qubit entanglement between ring-resonator photon-pair sources on a silicon chip}},\ }\href@noop {} {\bibfield  {journal} {\bibinfo  {journal} {Nat. Commun.}\ }\textbf {\bibinfo {volume} {6}} (\bibinfo {year} {2015})}\BibitemShut {NoStop}%
\bibitem [{\citenamefont {Savanier}\ \emph {et~al.}(2015)\citenamefont {Savanier}, \citenamefont {Kumar},\ and\ \citenamefont {Mookherjea}}]{savanier2}%
  \BibitemOpen
  \bibfield  {author} {\bibinfo {author} {\bibfnamefont {M.}~\bibnamefont {Savanier}}, \bibinfo {author} {\bibfnamefont {R.}~\bibnamefont {Kumar}},\ and\ \bibinfo {author} {\bibfnamefont {S.}~\bibnamefont {Mookherjea}},\ }\bibfield  {title} {\bibinfo {title} {{Optimizing photon-pair generation electronically using a p-i-n diode incorporated in a silicon microring resonator}},\ }\href@noop {} {\bibfield  {journal} {\bibinfo  {journal} {Appl. Phys. Lett.}\ }\textbf {\bibinfo {volume} {107}} (\bibinfo {year} {2015})}\BibitemShut {NoStop}%
\bibitem [{\citenamefont {Bonneau}\ \emph {et~al.}(2016)\citenamefont {Bonneau}, \citenamefont {Silverstone},\ and\ \citenamefont {Thompson}}]{bonneau}%
  \BibitemOpen
  \bibfield  {author} {\bibinfo {author} {\bibfnamefont {D.}~\bibnamefont {Bonneau}}, \bibinfo {author} {\bibfnamefont {J.}~\bibnamefont {Silverstone}},\ and\ \bibinfo {author} {\bibfnamefont {M.}~\bibnamefont {Thompson}},\ }\bibinfo {title} {{Silicon Photonics III Systems and Applications, Chapter 2 Silicon Quantum Photonics}}\ (\bibinfo  {publisher} {Springer},\ \bibinfo {year} {2016})\ pp.\ \bibinfo {pages} {41--82}\BibitemShut {NoStop}%
\bibitem [{\citenamefont {Agrawal}(2013)}]{agrawal}%
  \BibitemOpen
  \bibfield  {author} {\bibinfo {author} {\bibfnamefont {G.~P.}\ \bibnamefont {Agrawal}},\ }\bibinfo {title} {{Nonlinear Fiber Optics 5th ed.}}\ (\bibinfo  {publisher} {Academic Press},\ \bibinfo {year} {2013})\ pp.\ \bibinfo {pages} {397--400}\BibitemShut {NoStop}%
\bibitem [{\citenamefont {Savanier}\ and\ \citenamefont {Mookherjea}(2016)}]{savanier}%
  \BibitemOpen
  \bibfield  {author} {\bibinfo {author} {\bibfnamefont {R.}~\bibnamefont {Savanier}, \bibfnamefont {M.~Kumar}}\ and\ \bibinfo {author} {\bibfnamefont {S.}~\bibnamefont {Mookherjea}},\ }\bibfield  {title} {\bibinfo {title} {{Photon pair generation from compact silicon microring resonators using microwatt-level pump powers}},\ }\href@noop {} {\bibfield  {journal} {\bibinfo  {journal} {Opt. Express}\ }\textbf {\bibinfo {volume} {24}} (\bibinfo {year} {2016})}\BibitemShut {NoStop}%
\bibitem [{\citenamefont {Jiang}\ \emph {et~al.}(2015)\citenamefont {Jiang} \emph {et~al.}}]{jiang}%
  \BibitemOpen
  \bibfield  {author} {\bibinfo {author} {\bibfnamefont {W.}~\bibnamefont {Jiang}} \emph {et~al.},\ }\bibfield  {title} {\bibinfo {title} {{Silicon-chip source of bright photon pairs}},\ }\href@noop {} {\bibfield  {journal} {\bibinfo  {journal} {Opt. Express}\ }\textbf {\bibinfo {volume} {23}} (\bibinfo {year} {2015})}\BibitemShut {NoStop}%
\bibitem [{\citenamefont {Li}\ \emph {et~al.}(2025)\citenamefont {Li} \emph {et~al.}}]{bohan}%
  \BibitemOpen
  \bibfield  {author} {\bibinfo {author} {\bibfnamefont {B.}~\bibnamefont {Li}} \emph {et~al.},\ }\bibfield  {title} {\bibinfo {title} {{Down-converted photon pairs in a high-Q silicon nitride microresonator}},\ }\href@noop {} {\bibfield  {journal} {\bibinfo  {journal} {Nature}\ }\textbf {\bibinfo {volume} {639}} (\bibinfo {year} {2025})}\BibitemShut {NoStop}%
\bibitem [{\citenamefont {Meyer-Scott}\ \emph {et~al.}(2018)\citenamefont {Meyer-Scott} \emph {et~al.}}]{klyshko1}%
  \BibitemOpen
  \bibfield  {author} {\bibinfo {author} {\bibfnamefont {E.}~\bibnamefont {Meyer-Scott}} \emph {et~al.},\ }\bibfield  {title} {\bibinfo {title} {{High-performance source of spectrally pure, polarization entangled photon pairs based on hybrid integrated-bulk optics}},\ }\href@noop {} {\bibfield  {journal} {\bibinfo  {journal} {Opt. Express}\ }\textbf {\bibinfo {volume} {26}} (\bibinfo {year} {2018})}\BibitemShut {NoStop}%
\bibitem [{\citenamefont {Xiyuan}\ \emph {et~al.}(2016)\citenamefont {Xiyuan} \emph {et~al.}}]{klyshko2}%
  \BibitemOpen
  \bibfield  {author} {\bibinfo {author} {\bibfnamefont {L.}~\bibnamefont {Xiyuan}} \emph {et~al.},\ }\bibfield  {title} {\bibinfo {title} {{Heralding single photons from a high-Q silicon microdisk}},\ }\href@noop {} {\bibfield  {journal} {\bibinfo  {journal} {Optica}\ }\textbf {\bibinfo {volume} {3}} (\bibinfo {year} {2016})}\BibitemShut {NoStop}%
\bibitem [{\citenamefont {Silverstone}\ \emph {et~al.}(2014)\citenamefont {Silverstone} \emph {et~al.}}]{silverstone3}%
  \BibitemOpen
  \bibfield  {author} {\bibinfo {author} {\bibfnamefont {J.}~\bibnamefont {Silverstone}} \emph {et~al.},\ }\bibfield  {title} {\bibinfo {title} {{On-chip quantum interference between silicon photon-pair sources}},\ }\href@noop {} {\bibfield  {journal} {\bibinfo  {journal} {Nat. Photon.}\ }\textbf {\bibinfo {volume} {8}} (\bibinfo {year} {2014})}\BibitemShut {NoStop}%
\bibitem [{\citenamefont {Burridge}\ \emph {et~al.}(2022)\citenamefont {Burridge} \emph {et~al.}}]{benarxiv}%
  \BibitemOpen
  \bibfield  {author} {\bibinfo {author} {\bibfnamefont {B.}~\bibnamefont {Burridge}} \emph {et~al.},\ }\bibfield  {title} {\bibinfo {title} {{Quantifying Hidden Nonlinear Noise in Integrated Photonics}},\ }\href@noop {} {\bibfield  {journal} {\bibinfo  {journal} {arXiv:2209.14317 [quant-ph]}\ } (\bibinfo {year} {2022})}\BibitemShut {NoStop}%
\bibitem [{\citenamefont {Martini}\ and\ \citenamefont {Politi}(2018)}]{martini}%
  \BibitemOpen
  \bibfield  {author} {\bibinfo {author} {\bibfnamefont {F.}~\bibnamefont {Martini}}\ and\ \bibinfo {author} {\bibfnamefont {A.}~\bibnamefont {Politi}},\ }\bibfield  {title} {\bibinfo {title} {{Four wave mixing in 3C SiC ring resonators}},\ }\href@noop {} {\bibfield  {journal} {\bibinfo  {journal} {Appl. Phys. Lett.}\ }\textbf {\bibinfo {volume} {112}} (\bibinfo {year} {2018})}\BibitemShut {NoStop}%
\bibitem [{\citenamefont {Wang}\ \emph {et~al.}(2025)\citenamefont {Wang} \emph {et~al.}}]{wang}%
  \BibitemOpen
  \bibfield  {author} {\bibinfo {author} {\bibfnamefont {H.}~\bibnamefont {Wang}} \emph {et~al.},\ }\bibfield  {title} {\bibinfo {title} {{Bright Heralded Single-Photon Source Saturating Theoretical Single-photon Purity}},\ }\href@noop {} {\bibfield  {journal} {\bibinfo  {journal} {Laser Photonics Rev.}\ }\textbf {\bibinfo {volume} {19}} (\bibinfo {year} {2025})}\BibitemShut {NoStop}%
\bibitem [{\citenamefont {Rahmouni}\ \emph {et~al.}(2024)\citenamefont {Rahmouni} \emph {et~al.}}]{rahmouni}%
  \BibitemOpen
  \bibfield  {author} {\bibinfo {author} {\bibfnamefont {A.}~\bibnamefont {Rahmouni}} \emph {et~al.},\ }\bibfield  {title} {\bibinfo {title} {{Entangled photon pair generation in an integrated SiC platform}},\ }\href@noop {} {\bibfield  {journal} {\bibinfo  {journal} {Light Sci. Appl.}\ }\textbf {\bibinfo {volume} {13}} (\bibinfo {year} {2024})}\BibitemShut {NoStop}%
\bibitem [{\citenamefont {Zhu}\ \emph {et~al.}(2024)\citenamefont {Zhu} \emph {et~al.}}]{zhu2}%
  \BibitemOpen
  \bibfield  {author} {\bibinfo {author} {\bibfnamefont {X.}~\bibnamefont {Zhu}} \emph {et~al.},\ }\bibfield  {title} {\bibinfo {title} {{Enhanced efficiency of correlated photon pairs generation in silicon nitride with a low-loss 3D edge coupler}},\ }\href@noop {} {\bibfield  {journal} {\bibinfo  {journal} {APL Photon.}\ }\textbf {\bibinfo {volume} {9}} (\bibinfo {year} {2024})}\BibitemShut {NoStop}%
\bibitem [{\citenamefont {Chen}\ \emph {et~al.}(2024)\citenamefont {Chen} \emph {et~al.}}]{chen}%
  \BibitemOpen
  \bibfield  {author} {\bibinfo {author} {\bibfnamefont {R.}~\bibnamefont {Chen}} \emph {et~al.},\ }\bibfield  {title} {\bibinfo {title} {{Ultralow-Loss Integrated Photonics Enables Bright, Narrowband, Photon-Pair Sources}},\ }\href@noop {} {\bibfield  {journal} {\bibinfo  {journal} {Phys. Rev. Lett.}\ }\textbf {\bibinfo {volume} {133}} (\bibinfo {year} {2024})}\BibitemShut {NoStop}%
\bibitem [{\citenamefont {Mahmudlu}\ \emph {et~al.}(2021)\citenamefont {Mahmudlu} \emph {et~al.}}]{mahmudlu}%
  \BibitemOpen
  \bibfield  {author} {\bibinfo {author} {\bibfnamefont {H.}~\bibnamefont {Mahmudlu}} \emph {et~al.},\ }\bibfield  {title} {\bibinfo {title} {{AlGaAs-on-insulator waveguide for highly efficient photon-pair generation via spontaneous four-wave mixing}},\ }\href@noop {} {\bibfield  {journal} {\bibinfo  {journal} {Opt. Lett.}\ }\textbf {\bibinfo {volume} {46}} (\bibinfo {year} {2021})}\BibitemShut {NoStop}%
\bibitem [{\citenamefont {Choi}\ \emph {et~al.}(2020)\citenamefont {Choi} \emph {et~al.}}]{choi}%
  \BibitemOpen
  \bibfield  {author} {\bibinfo {author} {\bibfnamefont {J.}~\bibnamefont {Choi}} \emph {et~al.},\ }\bibfield  {title} {\bibinfo {title} {{Correlated photon pair generation in ultra-silicon-rich nitride waveguide}},\ }\href@noop {} {\bibfield  {journal} {\bibinfo  {journal} {Opt. Commun.}\ }\textbf {\bibinfo {volume} {463}} (\bibinfo {year} {2020})}\BibitemShut {NoStop}%
\bibitem [{\citenamefont {Faruque}\ \emph {et~al.}(2018{\natexlab{b}})\citenamefont {Faruque} \emph {et~al.}}]{faruque1}%
  \BibitemOpen
  \bibfield  {author} {\bibinfo {author} {\bibfnamefont {I.}~\bibnamefont {Faruque}} \emph {et~al.},\ }\bibfield  {title} {\bibinfo {title} {{On-chip quantum interference with heralded photons from two independent micro-ring resonator sources in silicon photonics}},\ }\href@noop {} {\bibfield  {journal} {\bibinfo  {journal} {Opt. Express}\ }\textbf {\bibinfo {volume} {26}} (\bibinfo {year} {2018}{\natexlab{b}})}\BibitemShut {NoStop}%
\bibitem [{\citenamefont {Ma}\ \emph {et~al.}(2018)\citenamefont {Ma}, \citenamefont {Wang},\ and\ \citenamefont {Mookherjea}}]{ma2}%
  \BibitemOpen
  \bibfield  {author} {\bibinfo {author} {\bibfnamefont {C.}~\bibnamefont {Ma}}, \bibinfo {author} {\bibfnamefont {X.}~\bibnamefont {Wang}},\ and\ \bibinfo {author} {\bibfnamefont {S.}~\bibnamefont {Mookherjea}},\ }\bibfield  {title} {\bibinfo {title} {{Photon-pair and heralded single photon generation initiated by a fraction of a 10 Gbps data stream}},\ }\href@noop {} {\bibfield  {journal} {\bibinfo  {journal} {Opt. Express}\ }\textbf {\bibinfo {volume} {26}} (\bibinfo {year} {2018})}\BibitemShut {NoStop}%
\bibitem [{\citenamefont {Mazeas}\ \emph {et~al.}(2016)\citenamefont {Mazeas} \emph {et~al.}}]{mazeas}%
  \BibitemOpen
  \bibfield  {author} {\bibinfo {author} {\bibfnamefont {F.}~\bibnamefont {Mazeas}} \emph {et~al.},\ }\bibfield  {title} {\bibinfo {title} {{High-quality photonic entanglement for wavelength-multiplexed quantum communication based on a silicon chip}},\ }\href@noop {} {\bibfield  {journal} {\bibinfo  {journal} {Opt. Express}\ }\textbf {\bibinfo {volume} {24}} (\bibinfo {year} {2016})}\BibitemShut {NoStop}%
\bibitem [{\citenamefont {Wakabayashi}\ \emph {et~al.}(2015)\citenamefont {Wakabayashi} \emph {et~al.}}]{wakabayashi}%
  \BibitemOpen
  \bibfield  {author} {\bibinfo {author} {\bibfnamefont {R.}~\bibnamefont {Wakabayashi}} \emph {et~al.},\ }\bibfield  {title} {\bibinfo {title} {{Time-bin entangled photon pair generation from Si micro-ring resonator}},\ }\href@noop {} {\bibfield  {journal} {\bibinfo  {journal} {Opt. Express}\ }\textbf {\bibinfo {volume} {23}} (\bibinfo {year} {2015})}\BibitemShut {NoStop}%
\end{thebibliography}%

\end{document}